\documentclass[aip,pop,reprint,amsmath,amssymb]{revtex4-1}
\pdfoutput=1
\usepackage[utf8]{inputenc}
\usepackage{color}
\usepackage{units}
\usepackage{mathrsfs}
\usepackage{mathtools}
\usepackage{graphicx}
\usepackage[unicode=true,
 bookmarks=true,bookmarksnumbered=false,bookmarksopen=false,
 breaklinks=false,pdfborder={0 0 0},backref=false,colorlinks=true]
 {hyperref}
\hypersetup{pdftitle={Obliquely propagating electromagnetic waves in magnetized kappa plasmas},
 pdfauthor={Rudi Gaelzer and Luiz F. Ziebell},
 pdfkeywords={Kappa plasmas; waves; dispersion relations; kinetic theory; 
 methods: analytical; hypergeometric and Meijer G  functions.},
 linkcolor=blue, citecolor=blue}
\makeatletter

\allowdisplaybreaks

\makeatother

\begin{document}
\global\long\def\pFq#1#2{\prescript{\vphantom{#1}}{#1}{F}_{#2}^{\vphantom{#2}}}

\title{Obliquely propagating electromagnetic waves in magnetized kappa plasmas}

\author{R. Gaelzer}

\email{rudi.gaelzer@ufrgs.br}

\author{L. F. Ziebell}

\email{luiz.ziebell@ufrgs.br}

\affiliation{Instituto de Física, UFRGS, 91501-970 Porto Alegre, RS, Brazil}
\begin{abstract}
Velocity distribution functions (VDFs) that exhibit a power-law dependence
on the high-energy tail have been the subject of intense research
by the plasma physics community. Such functions, known as kappa or
superthermal distributions, have been found to provide a better fitting
to the VDFs measured by spacecraft in the solar wind. One of the problems
that is being addressed on this new light is the temperature anisotropy
of solar wind protons and electrons. In the literature, the general
treatment for waves excited by (bi-)Maxwellian plasmas is well-established.
However, for kappa distributions, the wave characteristics have been
studied mostly for the limiting cases of purely parallel or perpendicular
propagation, relative to the ambient magnetic field. Contributions
to the general case of obliquely-propagating electromagnetic waves
have been scarcely reported so far. The absence of a general treatment
prevents a complete analysis of the wave-particle interaction in kappa
plasmas, since some instabilities can operate simultaneously both
in the parallel and oblique directions. In a recent work, Gaelzer
and Ziebell {[}J. Geophys. Res. \textbf{119}, 9334 (2014){]} obtained
expressions for the dielectric tensor and dispersion relations for
the low-frequency, quasi-perpendicular dispersive Alfvén waves resulting
from a kappa VDF\@. In the present work, the formalism is generalized
for the general case of electrostatic and/or electromagnetic waves
propagating in a kappa plasma in any frequency range and for arbitrary
angles. An isotropic distribution is considered, but the methods used
here can be easily applied to more general anisotropic distributions,
such as the bi-kappa or product-bi-kappa. 
\end{abstract}

\keywords{Kappa plasmas; waves; dispersion relations; kinetic theory; methods:
analytical; hypergeometric and Meijer $G$ functions.}

\maketitle

\section{Introduction}

A significant effort has been made in recent years on the study of
the properties of plasmas composed by particles described by the so-called
superthermal or kappa velocity distribution functions ($\kappa$VDFs)\@.
These distributions distinguish themselves from the usual Maxwell-Bolzmann
VDF by the presence of a tail (the high-velocity portion of the VDF)
that decays with velocity according to a power law, instead of the
Gaussian profile characteristic of the Maxwellian distribution.

Various space plasma environments, such as planetary magnetospheres,
the solar corona, or the solar wind, are composed by particles whose
observed VDFs are better fitted by a kappa or by combinations of kappa
distributions, instead of any possible combination of Maxwellians\@.\cite{Maksimovic+97/05,Zouganelis08/08,Stverak+09/05,Benson+13/08}
The morphological distinction between a Maxwellian VDF and a kappa
VDF is not a mere mathematical or observational curiosity. As a consequence
of this difference, the physical processes that occur inside these
environments are strongly influenced by the particular profile of
the distribution and can significantly differ from the behavior one
would expect from a quasi-thermal plasma.

Evidence of the importance of the particular morphology of superthermal
distributions has been appearing in the literature during the last
decade. Just to cite some examples, kappa distributions were employed
by Viñas \emph{et al.}\cite{Vinas+05/06} to provide a better description
of the plasma resonances observed in Earth's magnetosphere. Recent
studies concerning the quasi-thermal emission of magnetized plasmas
resulting from the single-particle fluctuations have revealed distinctive
differences whether the VDF is (bi-) Maxwellian or (bi-) kappa\@.\cite{SchlickeiserYoon12/02,Lazar+12/12,Felten+13/05,Vinas+14/01,Navarro+15/04,Vinas+15/05}
Simultaneously, several other studies have been conducted concerning
the wave-particle resonance in Maxwellian or kappa plasmas, both in
the low- and in the high-frequency regions of parallel-propagating
waves\@.\cite{LazarPoedts09/01,SeoughYoon09/09,Yoon+10/08,Lazar+11/01,Lazar+11/10,Lazar12/11,Lazar+14/01,LazarPoedts14/01,dosSantos+14/11,Lazar+15/01,dosSantos+15/12}
The last couple of research subjects are relevant to the problem of
the observed temperature anisotropy of both electronic and protonic
populations of the solar wind. In laboratory and tokamak plasmas,
kappa distributions have also been used to address discrepancies between
experiments and theory when Maxwellian VDFs are employed\@.\cite{Dumont+05/04}
Important problems such as cyclotron heating and wave resonance with
runaway (or superthermal) electrons have been considered by this approach\@.\cite{Zhou+12/03,Mirza+15/09}
Finally, it has been also observed that in a dusty plasma, the excess
of superthermal plasma particles affects not only the wave-resonance
characteristics (dispersion relations and damping/growth rates), but
alters the resulting electrical charge of the dust particles as well.\cite{Gaelzer+10/09,Deeba+10/10,Rubab+11/07,Galvao+11/12,Galvao+12/12,dosSantos+16}

Some theories have been proposed to address the origin of $\kappa$VDFs
from a fundamental set of postulates. The most accepted explanation
nowadays is based on the principle of nonadditive entropy proposed
by C. Tsallis\@.\cite{Tsallis09} According to Tsallis's postulate,
many-particle physical systems that evolve subjected to long-scale
correlations and nonlinear effects can reach a quasi-stationary (or
truly stationary) state in which the probability distribution functions
of the physical quantities (such as the VDF) are not given by a Boltzmann
distribution, but rather by probability functions that arise from
the maximization of the nonadditive $q$-entropy proposed by Tsallis.
A detailed account about the appearance of the $\kappa$VDF as a consequence
of the $q$-entropic principle and the resulting implications for
space (or laboratory) plasma physics was given by Livadiotis and McComas\@.\cite{LivadiotisMcComas13/05}
Some of the alternative or related theories proposed in recent years
are briefly mentioned in Ref. \onlinecite{GaelzerZiebell14/12}.

During the last generation, the yearly number of papers published
on the subject of kappa distributions has been growing by a measurable
exponential law\@.\cite{LivadiotisMcComas13/05} In spite of the
evident interest of the plasma community (mostly by space physicists)
on kappa distributions, the overwhelming majority of papers concerned
with the propagation and absorption/amplification of electrostatic/electromagnetic
waves in superthermal magnetized plasmas has been restricted to the
particular case of parallel propagation (relative to the ambient magnetic
field).\footnote{Some relevant references were given above. The Reader is once again
referred to Ref. \onlinecite{GaelzerZiebell14/12} for an additional
list of original and revision papers.}

The more general case of obliquely-propagating waves has been tackled
by relatively few works so far. A basic formalism was proposed by
Summers \emph{et al}\@.\cite{Summers+94/06} In their work, however,
the results were mostly obtained by numerical integrations. A similar
approach was later adopted by Basu\cite{Basu09/05} and Liu \emph{et
al}\@.\cite{Liu+14/03} Important as these works are, the evaluation
of the dielectric tensor components via numerical quadratures hampers
the derivation of analytical expressions for the dispersion relations
and damping/growth rates. Moreover, the analytical results obtained
by the above authors were derived from power series expansions that
do not account for all the possible mathematical properties of the
dielectric tensor, as will be demonstrated below.

Recently, Gaelzer and Ziebell\cite{GaelzerZiebell14/12} proposed
for the first time a mathematical formulation for dispersive Alfvén
waves (DAWs), which are low-frequency, quasi-perpendicular waves propagating,
in this case, in an isotropic kappa plasma. The proposed formulation
obtains analytical expressions for the dielectric tensor components
in terms of closed-form special functions that describe both the wave-particle
resonance and the Larmor radius effects due to a superthermal VDF
on the characteristics of the dispersive Alfvén waves. The formalism
that was introduced by Ref. \onlinecite{GaelzerZiebell14/12} rendered
the problem of wave propagation more tractable, and new expressions
for the dispersion relations of DAWs were in consequence obtained.

More recently, Astfalk \emph{et al}\@.\cite{Astfalk+15/09} reported
the numerical implementation of the method proposed by Ref. \onlinecite{Summers+94/06}
to study wave propagation at arbitrary angles in anisotropic superthermal
plasmas described by a bi-kappa VDF\@.

Even more recent, Sugiyama \emph{et al}\@.\cite{Sugiyama+15/10}
implemented a dispersion equation solver for electromagnetic ion-cyclotron
waves propagating at arbitrary angles in a kappa-Maxwellian plasma.
From the purely mathematical point of view, the treatment of kappa-Maxwellian
VDFs is simpler than isotropic kappa distributions, since in the former
the dependence of the distribution function on the parallel component
of the particle velocity (given by a $\kappa$VDF) factors from the
dependence on the perpendicular component, which is Maxwellian. Such
simplification is not possible with an isotropic $\kappa$VDF.

In this work, the initial formulation proposed by Gaelzer and Ziebell\cite{GaelzerZiebell14/12}
is generalized for any number of particle species, wave frequency
range and propagation angle. A thorough and comprehensive analysis
is conducted on the mathematical aspects and properties of the dielectric
tensor components for an isotropic kappa distribution function, which
are described by analytical, closed-form expressions. The said components
are given by special functions for a kappa plasma that are either
generalizations of already known expressions or are completely new
definitions. As a demonstration on the feasibility of the proposed
formalism, the dispersion equation is solved for high-frequency waves
propagating in various angles and for several values of the $\kappa$
parameter.

This paper is organized as follows. In section \ref{sec:Theoretical-base},
the velocity distribution function and the dispersion equation for
waves in a kappa plasma are considered. The dielectric tensor components
are written in terms of (thermal) Stix parameters. Section \ref{sec:Kappa-functions}
contains the bulk of the formalism. In there, the special functions
that appear in the dielectric tensor for a kappa plasma are discussed
in details. Section \ref{sec:Numerical-applications} describes a
simple implementation of the formalism presented in the previous sections.
Finally, section \ref{sec:Conclusions} contains the conclusions.
Additional and supporting matériel are provided in appendices \ref{sec:Additional_props-kappa_plasma_funcs},
\ref{sec:Hypergeometric-functions} and \ref{sec:DT_Cartesian}.

\section{Theoretical formulation\label{sec:Theoretical-base}}

\subsection{The velocity distribution function}

Calling $f_{a}\left(\boldsymbol{v}\right)$ the VDF for the plasma
species $a$, we will adopt in this work the isotropic kappa VDF form
already introduced by Ref. \onlinecite{GaelzerZiebell14/12}, 
\begin{equation}
f_{\kappa,a}\left(\boldsymbol{v}\right)=\frac{1}{\pi^{3/2}w_{a}^{3}}\frac{\kappa_{a}^{-3/2}\Gamma\left(\sigma_{a}\right)}{\Gamma\left(\sigma_{a}-\nicefrac{3}{2}\right)}\left(1+\frac{v^{2}}{\kappa_{a}w_{a}^{2}}\right)^{-\sigma_{a}},\label{eq:Kappa-VDF}
\end{equation}
which is valid when $\sigma_{a}>\nicefrac{3}{2}$\@. In (\ref{eq:Kappa-VDF}),
$\boldsymbol{v}$ is the particle's velocity, $\sigma_{a}=\kappa_{a}+\alpha_{a}$,
where $\kappa_{a}$ is the kappa index for the $a$-th species and
$\alpha_{a}$ is a free real parameter, $w_{a}=w_{a}\left(\kappa_{a}\right)$
is another parameter with the same physical dimension and meaning
as the particle's thermal speed and which depends on the $\kappa$
parameter. Finally, $\Gamma\left(z\right)$ is the gamma function.

For all practical purposes, it is assumed in this work the $\kappa$VDF
form first proposed by Summers and Thorne,\cite{SummersThorne91/08}
which can be obtained from (\ref{eq:Kappa-VDF}) by setting $\alpha_{a}=1$
and $w_{a}^{2}=\left(1-3/2\kappa_{a}\right)v_{Ta}^{2}$, where $v_{Ta}^{2}=2T_{a}/m_{a}$
is the thermal speed of species $a$ with mass $m_{a}$ and temperature
$T_{a}$ (in energy dimension)\@. For any application of the general
formalism presented in this work, which assumes an isotropic kappa
distribution for a given species, the general form given in Eq. (\ref{eq:Kappa-VDF})
reduces to the particular form introduced by Summers and Thorne (hereafter
called the ST91 model) with the choices just given. Nevertheless,
the index $\alpha_{a}$ will be kept throughout this work because
it can be useful in order to extend the formalism for anisotropic
$\kappa$VDFs, such as the bi-kappa\cite{SummersThorne92/11} or product-bi-kappa\cite{SummersThorne92/11,Livadiotis15/03}
models. The specific value of the $\alpha$-parameter may then depend
on the dimensionality of the distribution and on the correlation between
the different degrees of freedom of the plasma particles.\cite{Livadiotis15/03}

The distribution (\ref{eq:Kappa-VDF}) contains the expected limiting
form of the Maxwell-Boltzmann distribution, $f_{M,a}\left(\boldsymbol{v}\right)=\pi^{-3/2}v_{Ta}^{-3}\exp\left(-v^{2}/v_{Ta}^{2}\right)$,
when $\kappa_{a}\to\infty$\@. This property is easily verified by
using the exponential limit 
\begin{equation}
\lim_{\kappa\to\infty}\left(1+\frac{y^{2}}{\kappa}\right)^{-\kappa}=e^{-y^{2}}\label{eq:Exp-limit}
\end{equation}
and Stirling's formula\@.\cite{AskeyRoy-NIST10} This process of
limit evaluation will be dubbed here as the \emph{Maxwellian limit}.

The ST91 form has the additional property that the \emph{kinetic temperature}
of the $a$-th species, defined from the second moment of the distribution
by $T_{K,a}=\frac{1}{3}m_{a}\left\langle v^{2}\right\rangle =\frac{1}{3}m_{a}\int d^{3}v\,v^{2}f_{\kappa,a}\left(\boldsymbol{v}\right)$,
equals the thermodynamic measure of temperature, \emph{i. e.}, $T_{K,a}=T_{a}$\@.\cite{LivadiotisMcComas13/05,Livadiotis15/03}
This is an important property, since a single macroscopic parameter,
namely $T_{a}$, can be used to measure the velocity spread of plasma
species, independent on the particular value of the $\kappa$ parameter
$\left(\nicefrac{1}{2}<\kappa<\infty\right)$\@. Moreover, the ST91
form is also the most probable velocity distribution function for
a kappa plasma, as obtained from Tsallis's entropic principle\@.\cite{LivadiotisMcComas13/05,Livadiotis15/03}
This is another important property, since it establishes a theoretical
background with nonequilibrium statistical mechanics. These are the
main reasons why the ST91 form is by far the most frequently employed
to describe kappa plasmas.

\subsection{The dispersion equation}

Starting from the well-known expression for the dielectric tensor
of a homogeneous magnetized plasma,\cite{Brambilla98} 
\begin{multline}
\varepsilon_{ij}\left(\boldsymbol{k},\omega\right)=\delta_{ij}\\
+\sum_{a}\frac{\omega_{pa}^{2}}{\omega^{2}}\left[\sum_{n\to-\infty}^{\infty}\int d^{3}v\frac{v_{\perp}\left(\Xi_{na}\right)_{i}\left(\Xi_{na}^{*}\right)_{j}\mathcal{L}f_{a}}{\omega-n\Omega_{a}-k_{\parallel}v_{\parallel}}\right.\\
\left.+\delta_{iz}\delta_{jz}\int d^{3}v\frac{v_{\parallel}}{v_{\perp}}Lf_{a}\right],\label{eq:DT_Cartesian}
\end{multline}
where $\left\{ i,j\right\} =\left\{ x,y,z\right\} $ identifies the
Cartesian (in the $E^{3}$ space) components of $\varepsilon_{ij}$,
with $\left\{ \hat{\boldsymbol{x}},\hat{\boldsymbol{y}},\hat{\boldsymbol{z}}\right\} $
being the basis vectors of $E^{3}$, $\boldsymbol{\Xi}_{na}=n\rho_{a}^{-1}J_{n}\left(\rho_{a}\right)\hat{\boldsymbol{x}}-iJ_{n}'\left(\rho_{a}\right)\hat{\boldsymbol{y}}+\left(v_{\parallel}/v_{\perp}\right)J_{n}\left(\rho_{a}\right)\hat{\boldsymbol{z}}$,
where $J_{n}\left(z\right)$ is the Bessel function of the first kind,\cite{OlverMaximon-NIST10}
$\rho_{a}=k_{\perp}v_{\perp}/\Omega_{a}$, $Lf_{a}=v_{\perp}\partial f_{a}/\partial v_{\parallel}-v_{\parallel}\partial f_{a}/\partial v_{\perp}$,
and $\mathcal{L}f_{a}=\omega\partial f_{a}/\partial v_{\perp}+k_{\parallel}Lf_{a}$\@.
Also, $\omega_{pa}^{2}=4\pi n_{a}q_{a}^{2}/m_{a}$ and $\Omega_{a}=q_{a}B_{0}/m_{a}c$
are the plasma and cyclotron frequencies of the $a$-th species, respectively,
$\omega$ and $\boldsymbol{k}=k_{\perp}\hat{\boldsymbol{x}}+k_{\parallel}\hat{\boldsymbol{z}}$
are the wave frequency and wavenumber, $\boldsymbol{B}_{0}=B_{0}\hat{\boldsymbol{z}}$
$\left(B_{0}>0\right)$ is the ambient magnetic induction field and
the symbols $\parallel\left(\perp\right)$ denote the usual parallel
(perpendicular) components of vectors/tensors, respective to $\boldsymbol{B}_{0}$\@.

The wave equation in Fourier space can be written as $\Lambda_{ij}\left(\boldsymbol{k},\omega\right)E_{j}\left(\boldsymbol{k},\omega\right)=0$,
where the Einstein convention of implicit sum over repeated indices
is adopted, $\Lambda_{ij}\left(\boldsymbol{k},\omega\right)=N_{i}N_{j}-N^{2}\delta_{ij}+\varepsilon_{ij}\left(\boldsymbol{k},\omega\right)$
is the dispersion tensor, and where $\boldsymbol{N}=\boldsymbol{k}c/\omega$
is the refractive index. Finally, the dispersion relations are the
solutions of the dispersion equation 
\begin{equation}
\Lambda\left(\boldsymbol{k},\omega\right)=\det\left(\Lambda_{ij}\right)=0.\label{eq:DE_general}
\end{equation}

There are several known approximations and different expressions for
the dispersion equation, depending on physical parameters and propagation
characteristics, such as wave frequency range, propagation angle and
plasma species. Since this work develops a general formulation for
wave propagation in kappa plasmas, valid for any such characteristics,
a general form for the dispersion equation will be employed which,
albeit possibly not the more adequate for a particular situation,
was nevertheless able to provide initial explicit results from the
formalism.

With this objective in mind, it was found more convenient to change
the reference frame from the Cartesian to a \emph{rotated frame},
in which the usual limiting expressions for parallel or perpendicular
propagation angles are readily identified. In the rotated frame, the
dielectric tensor components are given by\cite{Brambilla98} 
\begin{equation}
\begin{aligned}\varepsilon_{++} & =\frac{1}{2}\left(\varepsilon_{xx}+\varepsilon_{yy}\right)-i\varepsilon_{xy} & \varepsilon_{+-} & =\frac{1}{2}\left(\varepsilon_{xx}-\varepsilon_{yy}\right)\\
\varepsilon_{--} & =\frac{1}{2}\left(\varepsilon_{xx}+\varepsilon_{yy}\right)+i\varepsilon_{xy} & \varepsilon_{+\parallel} & =\frac{1}{\sqrt{2}}\left(\varepsilon_{xz}+i\varepsilon_{yz}\right)\\
\varepsilon_{\parallel\parallel} & =\varepsilon_{zz} & \varepsilon_{-\parallel} & =\frac{1}{\sqrt{2}}\left(\varepsilon_{xz}-i\varepsilon_{yz}\right),
\end{aligned}
\label{eq:DT_rotated}
\end{equation}
in terms of the Cartesian components.

Finally, the rotated components can be expressed in terms of the \emph{thermal
Stix parameters} $\hat{L}$, $\hat{R}$, $\hat{P}$, $\hat{\tau}$,
$\hat{\mu}$ and $\hat{\nu}$ as\cite{Brambilla98} 
\begin{align*}
\varepsilon_{++} & =\hat{L}-N_{-}^{2}\left(\hat{\tau}-1\right) & \varepsilon_{--} & =\hat{R}-N_{+}^{2}\left(\hat{\tau}-1\right)\\
\varepsilon_{\parallel\parallel} & =\hat{P} & \varepsilon_{+-} & =N_{+}N_{-}\left(\hat{\tau}-1\right)\\
\varepsilon_{+\parallel} & =N_{+}N_{\parallel}\left(\hat{\mu}-1\right) & \varepsilon_{-\parallel} & =N_{-}N_{\parallel}\left(\hat{\nu}-1\right),
\end{align*}
where the first three parameters are, respectively, the usual $\hat{L}$eft,
$\hat{R}$ight and $\hat{P}$lasma Stix parameters (with thermal corrections),
which still exist in the cold plasma limit, whereas the last three
are only present when there are thermal (kinetic) effects.\footnote{Notice that the definitions adopted here for these last parameters
are slightly different from the definitions in Ref. \onlinecite{Brambilla98}.}

Inserting now the kappa distribution function (\ref{eq:Kappa-VDF})
into Eqs. (\ref{eq:DT_Cartesian}) and (\ref{eq:DT_rotated}), one
obtains, after a fair amount of algebra, the following compact expressions
for the (kappa) Stix parameters,\begin{subequations}\label{eq:Stix_pars_kappa}
\begin{align}
{\hat{L}_{\kappa} \choose \hat{R}_{\kappa}} & =1+\sum_{a}\frac{\omega_{pa}^{2}}{\omega^{2}}\xi_{0a}\nonumber \\
 & \times\sum_{n\to-\infty}^{\infty}n\left(\frac{n}{\mu_{a}}\pm\frac{\partial\hphantom{\mu_{a}}}{\partial\mu_{a}}\right)\mathcal{Z}_{n,\kappa_{a}}^{\left(\alpha_{a},2\right)}\\
\hat{P}_{\kappa} & =1-\sum_{a}\frac{\omega_{pa}^{2}}{\omega^{2}}\xi_{0a}\sum_{n\to-\infty}^{\infty}\xi_{na}\frac{\partial\hphantom{\xi_{na}}}{\partial\xi_{na}}\mathcal{Z}_{n,\kappa_{a}}^{\left(\alpha_{a},1\right)}\\
\hat{\tau}_{\kappa} & =1+\sum_{a}\frac{\omega_{pa}^{2}}{\Omega_{a}^{2}}\frac{w_{a}^{2}}{c^{2}}\xi_{0a}\sum_{n\to-\infty}^{\infty}\mathcal{Y}_{n,\kappa_{a}}^{\left(\alpha_{a},2\right)}\\
{\hat{\mu}_{\kappa} \choose \hat{\nu}_{\kappa}} & =1-\frac{1}{2}\sum_{a}\frac{\omega_{pa}^{2}}{\omega\Omega_{a}}\frac{w_{a}^{2}}{c^{2}}\xi_{0a}^{2}\nonumber \\
 & \times\sum_{n\to-\infty}^{\infty}\frac{\partial\hphantom{\xi_{na}}}{\partial\xi_{na}}\left(\frac{n}{\mu_{a}}\pm\frac{\partial\hphantom{\mu_{a}}}{\partial\mu_{a}}\right)\mathcal{Z}_{n,\kappa_{a}}^{\left(\alpha_{a},1\right)},
\end{align}
\end{subequations}where $\mu_{a}=\nu_{a}^{2}/2$, $\nu_{a}=k_{\perp}w_{a}/\Omega_{a}$,
$\xi_{na}=\left(\omega-n\Omega_{a}\right)/k_{\parallel}w_{a}$, and
where $\mathcal{Z}_{n,\kappa}^{\left(\alpha,\beta\right)}\left(\mu,\xi\right)$
and $\mathcal{Y}_{n,\kappa}^{\left(\alpha,\beta\right)}\left(\mu,\xi\right)$
are special functions that will be defined \emph{a posteriori} in
section \ref{sub:2_vars-PF}\@. In the same section, it will be shown
that in the Maxwellian limit the functions $\mathcal{Z}$ and $\mathcal{Y}$
reduce to 
\begin{equation}
\begin{aligned}\lim_{\kappa\to\infty}\mathcal{Z}_{n,\kappa}^{\left(\alpha,\beta\right)}\left(\mu,\xi\right) & =\mathscr{H}_{n}\left(\mu\right)Z\left(\xi\right)\\
\lim_{\kappa\to\infty}\mathcal{Y}_{n,\kappa}^{\left(\alpha,\beta\right)}\left(\mu,\xi\right) & =\mathscr{H}_{n}'(\mu)Z\left(\xi\right),
\end{aligned}
\label{eq:ZY_cal-Maxwell}
\end{equation}
where $Z\left(\zeta\right)$ is the usual plasma dispersion (or Fried
\& Conte) function, given below by definition (\ref{eq:Fried-Conte_function}),
and 
\begin{equation}
\mathscr{H}_{n}\left(z\right)=e^{-z}I_{n}\left(z\right)\label{eq:Thermal_PGF}
\end{equation}
is called here the (Maxwellian) \emph{plasma gyroradius function},
with $I_{n}\left(z\right)$ being the modified Bessel function\@.\cite{OlverMaximon-NIST10}
The expressions for the Stix parameters given by (\ref{eq:Stix_pars_kappa})
in the Maxwellian limit reduce to the well-known expressions that
can be found, \emph{e. g.}, in Ref. \onlinecite{Brambilla98}.

It is worth mentioning here that with the definition of the special
functions $\mathcal{Z}_{n,\kappa}^{\left(\alpha,\beta\right)}$ and
$\mathcal{Y}_{n,\kappa}^{\left(\alpha,\beta\right)}$, the Stix parameters
for a kappa plasma can be expressed in a form as compact as the correspondent
Maxwellian expressions. It will be also shown in section \ref{sub:2_vars-PF}
that these functions can always be evaluated from analytical, closed-form
expressions that do not depend on any residual numerical integration.
Therefore, with the formulation developed in this work, the evaluation
of (\ref{eq:Stix_pars_kappa}a-d) can be accomplished in a computational
time-frame comparable to the usual Maxwellian limit, with no increased
overhead due to lengthy numerical quadratures. Additionaly, the analytical
closed-form expressions obtained here simplify the determination of
mathematical properties of the special functions that arise from the
kappa distribution (\ref{eq:Kappa-VDF}), which in turn allows the
derivation of adequate approximations to the dielectric tensor and
dispersion relations, relevant to a particular wave propagation regime.

Finally, using the Stix parameters given in Eqs. (\ref{eq:Stix_pars_kappa}a-d),
the dispersion equation (\ref{eq:DE_general}) can be cast in the
following form\cite{Brambilla98}\begin{subequations}\label{eq:DE_rotated}
\begin{equation}
H_{X}\left(\boldsymbol{k},\omega\right)H_{O}\left(\boldsymbol{k},\omega\right)+\frac{1}{2}N_{\perp}^{2}N_{\parallel}^{2}K\left(\boldsymbol{k},\omega\right)=0,
\end{equation}
where 
\begin{align}
H_{X}\left(\boldsymbol{k},\omega\right)= & \left(\hat{L}-N_{\parallel}^{2}\right)\left(\hat{R}-N_{\parallel}^{2}\right)\nonumber \\
 & -N_{\perp}^{2}\left[\frac{1}{2}\left(\hat{L}+\hat{R}\right)-N_{\parallel}^{2}\right]\hat{\tau}\\
H_{O}\left(\boldsymbol{k},\omega\right)= & \hat{P}-N_{\perp}^{2}\\
K\left(\boldsymbol{k},\omega\right)= & \frac{1}{2}N_{\perp}^{2}\left(\hat{\nu}+\hat{\mu}\right)^{2}\hat{\tau}-\left(\hat{L}-N_{\parallel}^{2}\right)\hat{\nu}^{2}\nonumber \\
 & -\left(\hat{R}-N_{\parallel}^{2}\right)\hat{\mu}^{2}.
\end{align}
\end{subequations}

Equation (\ref{eq:DE_rotated}a) is convenient because it reduces
to simple forms for limiting propagation angles. For parallel propagation
$\left(k_{\perp}=0\right)$, it factors into the equations for the
left- and right-handed circularly polarized modes and the longitudinal
(or plasma) mode. On the other hand, for perpendicular propagation
$\left(k_{\parallel}=0\right)$, Eq. (\ref{eq:DE_rotated}a) factors
into the equations for the extraordinary and ordinary modes.

For the sake of completeness, and should the necessity arise, the
corresponding expressions for the Cartesian components of the dielectric
tensor for a kappa plasma are given in appendix \ref{sec:DT_Cartesian}.

In the next section, the definitions and mathematical properties of
the various special functions that appear in the treatment of a kappa
plasma are discussed.

\section{The special functions for a kappa plasma\label{sec:Kappa-functions}}

The Stix parameters for a kappa plasma, shown in Eqs. (\ref{eq:Stix_pars_kappa}a-d),
are given in terms of the two-variable special functions $\mathcal{Z}_{n,\kappa}^{\left(\alpha,\beta\right)}\left(\mu,\xi\right)$
and $\mathcal{Y}_{n,\kappa}^{\left(\alpha,\beta\right)}\left(\mu,\xi\right)$\@.
Before the proper definitions for these functions can be made, it
is necessary to give prior definitions and discuss the properties
of two related one-variable functions, namely the \emph{superthermal
plasma dispersion function }and the \emph{superthermal plasma gyroradius
function}. The most important properties will be shown in this section,
with additional properties given in appendix \ref{sec:Additional_props-kappa_plasma_funcs}.

\subsection{The superthermal plasma dispersion function}

This special function is the equivalent, for a kappa plasma, to the
well-known plasma dispersion (Fried \& Conte) function, defined by\cite{Brambilla98,FriedConte61}
\begin{equation}
Z(\xi)=\frac{1}{\sqrt{\pi}}\int_{-\infty}^{\infty}\frac{\mathrm{e}^{-y^{2}}dy}{y-\xi},\qquad\left(\text{for }\xi_{i}>0\right),\label{eq:Fried-Conte_function}
\end{equation}
where $\xi_{i}$ is the imaginary part of $\xi$.

The definition of the superthermal plasma dispersion function ($\kappa$PDF)
follows from the generalized expression adopted in this work for the
distribution function, Eq. (\ref{eq:Kappa-VDF}), and is given by
\begin{multline}
Z_{\kappa}^{\left(\alpha,\beta\right)}\left(\xi\right)=\frac{1}{\pi^{1/2}\kappa^{\beta+1/2}}\frac{\Gamma\left(\lambda-1\right)}{\Gamma\left(\sigma-\nicefrac{3}{2}\right)}\\
\times\int_{-\infty}^{\infty}ds\frac{\left(1+s^{2}/\kappa\right)^{-\left(\lambda-1\right)}}{s-\xi},\quad\left({\xi_{i}>0\atop \lambda>1}\right).\label{eq:Kappa-PDF}
\end{multline}
The function $Z_{\kappa}^{\left(\alpha,\beta\right)}\left(\xi\right)$
was first defined by Gaelzer and Ziebell\@.\cite{GaelzerZiebell14/12}
The parameter $\alpha$ is the same as it appears in (\ref{eq:Kappa-VDF}),
whereas $\beta$ is another real parameter. Moreover, $\lambda=\sigma+\beta$
$\left(\sigma=\kappa+\alpha\right)$\@. It is also noteworthy that
the definition (\ref{eq:Kappa-PDF}) is valid for $\xi_{i}>0$, as
is the definition of the Fried \& Conte function, but the integrand
of $Z_{\kappa}^{\left(\alpha,\beta\right)}\left(\xi\right)$ has additional
branch points at $s=\pm i\sqrt{\kappa}$, when $\lambda$ is noninteger\@.

Again, for an isotropic $\kappa$VDF, one can simply adopt the ST91
form, set $\alpha=1$ and erase all reference to the parameter $\alpha$\@.
However, the new parameter $\beta$ should be kept, because its value
is related to the wave polarization. For instance, using the ST91
form and setting $\beta=1$, Eq. (\ref{eq:Kappa-PDF}) reduces to
the original function $Z_{\kappa}^{*}\left(\xi\right)$ employed by
Summers and Thorne\cite{SummersThorne91/08} and Mace and Hellberg\cite{MaceHellberg95/06}
for a kappa (Lorentzian) plasma. It is interesting to mention that
the function $Z_{\kappa}^{*}\left(\xi\right)$ appears in the dispersion
equation for longitudinal (Langmuir, ion-sound) waves propagating
in a kappa plasma along the ambient magnetic field.

As another example, setting $\beta=0$ in (\ref{eq:Kappa-PDF}), one
obtains exactly the function $Z_{\kappa M}\left(\xi\right)$ defined
by Ref. \onlinecite{HellbergMace02/05} and related to parallel-propagating
electromagnetic waves in a kappa-Maxwellian plasma, or the function
$Z_{\kappa}^{0}\left(g\right)$ defined by Ref. \onlinecite{LazarPoedts09/01},
also related to circularly-polarized waves propagating in a kappa
plasma.

Hence, with the definition for the function $Z_{\kappa}^{\left(\alpha,\beta\right)}\left(\xi\right)$
given by Eq. (\ref{eq:Kappa-PDF}), one can describe the propagation
of different waves in a magnetized kappa plasma with a single expression,
in which case the parameter $\beta$ will be related to the wave polarization
and other characteristics, as will be shown below.

Several mathematical properties of $Z_{\kappa}^{\left(\alpha,\beta\right)}\left(\xi\right)$
will be derived in this section and in appendix \ref{sec:Additional_props-kappa_plasma_funcs}\@.
Some of the results shown here are generalizations of properties previously
obtained in various works found in the literature.\cite{SummersThorne91/08,SummersThorne92/11,MaceHellberg95/06,Mace96/06,HellbergMace02/05,Mace03/06,Mace04/02,MaceHellberg09/07}

\paragraph{Particular values. }

Direct integration of (\ref{eq:Kappa-PDF}) provides special values
of $Z_{\kappa}^{\left(\alpha,\beta\right)}\left(\xi\right)$ at special
points.

\paragraph{Value at $\xi=0$\@.}

At this point, 
\begin{multline*}
Z_{\kappa}^{\left(\alpha,\beta\right)}\left(0\right)=\frac{1}{\pi^{1/2}\kappa^{1/2+\beta}}\frac{\Gamma\left(\lambda-1\right)}{\Gamma\left(\sigma-\nicefrac{3}{2}\right)}\\
\times\int_{-\infty}^{\infty}ds\,s^{-1}\left(1+s^{2}/\kappa\right)^{-\left(\lambda-1\right)},
\end{multline*}
where the integration must be done following Landau prescription.
This means that it is possible to employ Plemelj formula to evaluate
the integral, which results equal to $i\pi$\@. Therefore, 
\begin{equation}
Z_{\kappa}^{\left(\alpha,\beta\right)}\left(0\right)=\frac{i\sqrt{\pi}\Gamma\left(\lambda-1\right)}{\kappa^{\beta+1/2}\Gamma\left(\sigma-\nicefrac{3}{2}\right)},\quad\left(\lambda>1\right).\label{eq:kPDF-origin}
\end{equation}
At the Maxwellian limit, one obtains $Z_{\kappa}^{\left(\alpha,\beta\right)}\left(0\right)\xrightarrow{\kappa\to\infty}i\sqrt{\pi}=Z\left(0\right)$
as expected.

\paragraph{Values at $\xi=\pm i\sqrt{\kappa}$\@.}

At these points, it can be shown that for $\lambda>1$, 
\[
\int_{-\infty}^{\infty}ds\frac{\left(1+s^{2}/\kappa\right)^{1-\lambda}}{s\mp i\sqrt{\kappa}}=\pm\frac{2i}{\sqrt{\kappa}}\int_{0}^{\infty}ds\left(1+\frac{s^{2}}{\kappa}\right)^{-\lambda}.
\]
Since the remaining integral is a special case of Euler's Beta integral,\cite{AskeyRoy-NIST10}
then 
\[
Z_{\kappa}^{\left(\alpha,\beta\right)}\left(\pm i\sqrt{\kappa}\right)=\pm i\frac{\kappa^{-\beta-1/2}\Gamma\left(\lambda-\nicefrac{1}{2}\right)}{\left(\lambda-1\right)\Gamma\left(\sigma-\nicefrac{3}{2}\right)}.
\]

\paragraph{Representations for $Z_{\kappa}^{\left(\alpha,\beta\right)}\left(\xi\right)$\@.}

The $\kappa$PDF has an already well-known representation in terms
of the Gauss hypergeometric function $\pFq 21\bigl({a,b\atop c};z\bigr)$,
defined in Eq. (\ref{eq:2F1_series})\@. This representation was
first derived by Ref. \onlinecite{MaceHellberg95/06} via an elegant
application of the residue theorem. An alternative and equivalent
derivation was already used by Refs. \onlinecite{Gaelzer+10/09,GaelzerZiebell14/12}
and will be employed again here.

First, by means of the variable transformation $t^{-1}=1+s^{2}/\kappa$,
the integral in (\ref{eq:Kappa-PDF}) becomes 
\begin{multline*}
Z_{\kappa}^{\left(\alpha,\beta\right)}\left(\xi\right)=\frac{\xi}{\pi^{1/2}\kappa^{\beta+1}}\frac{\Gamma\left(\lambda-1\right)}{\Gamma\left(\sigma-\nicefrac{3}{2}\right)}\\
\times\int_{0}^{1}t^{\lambda-3/2}\left(1-t\right)^{-1/2}\left[1-\left(1+\frac{\xi^{2}}{\kappa}\right)t\right]^{-1}dt.
\end{multline*}
Identifying with the integral representation (\ref{eq:2F1_integral_rep}),
one can write 
\begin{multline}
Z_{\kappa}^{\left(\alpha,\beta\right)}\left(\xi\right)=\frac{\kappa^{-\beta-1}\Gamma\left(\lambda-\nicefrac{1}{2}\right)}{\left(\lambda-1\right)\Gamma\left(\sigma-\nicefrac{3}{2}\right)}\xi\\
\times\pFq 21\left({1,\lambda-\frac{1}{2}\atop \lambda};1+\frac{\xi^{2}}{\kappa}\right),\;\left(\lambda>\frac{1}{2}\right).\label{eq:kPDF_rep1}
\end{multline}

Although a valid representation of $Z_{\kappa}^{\left(\alpha,\beta\right)}\left(\xi\right)$,
its principal branch is restricted to the sector $0<\arg\xi\leqslant\pi$
(\emph{i. e.}, to $\xi_{i}>0$)\@. Hence, this is not the adequate
representation when $\xi_{i}\leqslant0$, in other words, it does
not obey the Landau prescription. Nevertheless, if one employs the
analytic continuation formula (\ref{eq:2F1_analytic_cont}), a valid
representation is obtained, which is shown in Eq. (\ref{eq:kPDF_rep1c}).

For plasma physics applications, one is interested in a mathematical
representation of $Z_{\kappa}^{\left(\alpha,\beta\right)}\left(\xi\right)$
that satisfies the Landau prescription, \emph{i. e.}, is continuous
at the limit $\xi_{i}\to0$\@. Moreover, it is also required that
the sought representation lends itself to the numerical evaluation
of $Z_{\kappa}^{\left(\alpha,\beta\right)}\left(\xi\right)$, as carried
out by computer programming languages such as \texttt{Fortran}, \texttt{C/C++}
or \texttt{python}, and/or by computer algebra software. Such representations
are called in this work \emph{computable representations} of the $\kappa$PDF\emph{\@.
}Expression (\ref{eq:kPDF_rep1}) does not satisfy this requisite,
but it can be used within convenient transformations for the Gauss
function, thereby rendering other representations which are indeed
computable.

One such representation is obtained by inserting (\ref{eq:kPDF_rep1})
into the quadratic transformation (\ref{eq:2F1-transf_quadratic}),
resulting 
\begin{multline}
Z_{\kappa}^{\left(\alpha,\beta\right)}\left(\xi\right)=i\frac{\kappa^{-\beta-1/2}\Gamma\left(\lambda-\nicefrac{1}{2}\right)}{\left(\lambda-1\right)\Gamma\left(\sigma-\nicefrac{3}{2}\right)}\\
\times\pFq 21\left[{1,2\left(\lambda-1\right)\atop \lambda};\frac{1}{2}\left(1+\frac{i\xi}{\kappa^{1/2}}\right)\right].\label{eq:kPDF_rep2}
\end{multline}
Setting $\alpha=\beta=1$, representation (\ref{eq:kPDF_rep2}) of
$Z_{\kappa}^{\left(1,1\right)}\left(\xi\right)$ is exactly the original
result obtained by Ref. \onlinecite{MaceHellberg95/06}\@. On the
other hand, with $\alpha=1$ and $\beta=0$, there results again the
function $Z_{\kappa M}\left(\xi\right)$ obtained by Ref. \onlinecite{HellbergMace02/05}\@.
Hence, result (\ref{eq:kPDF_rep2}) generalizes these well-known representations.

It is important to mention at this point that although the form (\ref{eq:kPDF_rep2})
is indeed continuous across the real line of $\xi$, the branch cut
has not disappeared. It has just moved to the line $-\sqrt{\kappa}\geqslant\xi_{i}>-\infty$\@.
Therefore, it would be necessary to evaluate the analytical continuation
of (\ref{eq:kPDF_rep2}), should the branch line ever been crossed
during the dynamical evolution of the waves in the plasma. The same
caveat applies to all computable representations of $Z_{\kappa}^{\left(\alpha,\beta\right)}\left(\xi\right)$
found in this work.

Expression (\ref{eq:kPDF_rep2}) is well-suited for computational
purposes, however other forms are more convenient to derive further
analytical expression for $Z_{\kappa}^{\left(\alpha,\beta\right)}\left(\xi\right)$\@.
A representation that directly renders a series expansion for the
$\kappa$PDF is obtained by inserting now (\ref{eq:kPDF_rep1}) into
(\ref{eq:2F1-transf_linear-c}), resulting in 
\begin{gather}
Z_{\kappa}^{\left(\alpha,\beta\right)}\left(\xi\right)=-\frac{2\Gamma\left(\lambda-\nicefrac{1}{2}\right)\xi}{\kappa^{\beta+1}\Gamma\left(\sigma-\nicefrac{3}{2}\right)}\pFq 21\left({1,\lambda-\nicefrac{1}{2}\atop \nicefrac{3}{2}};-\frac{\xi^{2}}{\kappa}\right)\nonumber \\
+\frac{i\pi^{1/2}\Gamma\left(\lambda-1\right)}{\kappa^{\beta+1/2}\Gamma\left(\sigma-\nicefrac{3}{2}\right)}\left(1+\frac{\xi^{2}}{\kappa}\right)^{-\left(\lambda-1\right)},\label{eq:kPDF_rep3}
\end{gather}
also valid for $\lambda>\nicefrac{1}{2}$ and where identity (\ref{eq:1F0})
was also used. This form immediately renders the power series expansion,
given by (\ref{eq:kPDF-power_series}).

The representation (\ref{eq:kPDF_rep3}) is also important because
its Maxwellian limit reduces to a known representation of the Fried
\& Conte function. This fact can be verified by applying the limit
$\kappa\to\infty$ to (\ref{eq:kPDF_rep3}), employing both the exponential
limit (\ref{eq:Exp-limit}) and the Stirling formula, obtaining thus
\[
Z_{\kappa}^{\left(\alpha,\beta\right)}\left(\xi\right)\xrightarrow{\kappa\to\infty}-2\xi\pFq 11\left({1\atop \nicefrac{3}{2}};-\xi^{2}\right)+i\pi^{1/2}e^{-\xi^{2}}.
\]
The function $\pFq 11\bigl({1\atop \nicefrac{3}{2}};-\xi^{2}\bigr)$,
according to (\ref{eq:Kummer-1F1-series}), is a particular case of
the Kummer confluent hypergeometric function, and this result is another
known representation for the Fried \& Conte function.\cite{Peratt84/03}

The last representation for $Z_{\kappa}^{\left(\alpha,\beta\right)}\left(\xi\right)$
to be shown in this section is obtained from (\ref{eq:kPDF_rep3})
with the use of (\ref{eq:2F1-transf_linear-d}), resulting in 
\begin{multline}
Z_{\kappa}^{\left(\alpha,\beta\right)}\left(\xi\right)=-\frac{\Gamma\left(\lambda-\nicefrac{3}{2}\right)\xi^{-1}}{\kappa^{\beta}\Gamma\left(\sigma-\nicefrac{3}{2}\right)}\pFq 21\left({1,\nicefrac{1}{2}\atop \nicefrac{5}{2}-\lambda};-\frac{\kappa}{\xi^{2}}\right)\\
+\left[i-\tan\left(\lambda\pi\right)\right]\frac{\pi^{1/2}\Gamma\left(\lambda-1\right)}{\kappa^{\beta+1/2}\Gamma\left(\sigma-\nicefrac{3}{2}\right)}\left(1+\frac{\xi^{2}}{\kappa}\right)^{1-\lambda},\label{eq:kPDF_rep4}
\end{multline}
which is now valid for $\;\lambda>\nicefrac{3}{2}$ and $\lambda\neq m+\nicefrac{3}{2}$,
where $m=1,2,\dots$\@. This form is well-suited to obtain an asymptotic
approximation for $Z_{\kappa}^{\left(\alpha,\beta\right)}\left(\xi\right)$,
shown in appendix \ref{sec:Additional_props-kappa_plasma_funcs}\@.
However, now there are two additional restrictions. First, when $\lambda$
is semi-integer, each term has a singularity. Fortunately they cancel
out and in the appendix \ref{sec:Additional_props-kappa_plasma_funcs}
it is also shown the expression for (\ref{eq:kPDF_rep4}) when $\lambda=m+\nicefrac{3}{2}$\@.
The other restriction appears due to the fact that now the branch
line is stretched along the whole imaginary axis $-\infty<\xi_{i}<\infty$.

\paragraph{Recurrence relations.}

Some important recurrence relations for $Z_{\kappa}^{\left(\alpha,\beta\right)}\left(\xi\right)$
can be obtained. Performing integrations by parts in (\ref{eq:Kappa-PDF}),
one obtains\begin{subequations} 
\begin{align}
\left(1+\frac{\xi^{2}}{\kappa}\right)Z_{\kappa}^{\left(\alpha+1,\beta\right)}\left(\xi\right) & =\frac{\lambda-1}{\sigma-\nicefrac{3}{2}}Z_{\kappa}^{\left(\alpha,\beta\right)}\left(\xi\right)\nonumber \\
 & -\frac{\Gamma\left(\lambda-\nicefrac{1}{2}\right)}{\kappa^{\beta+1}\Gamma\left(\sigma-\nicefrac{1}{2}\right)}\xi,\\
\left(1+\frac{\xi^{2}}{\kappa}\right)Z_{\kappa}^{\left(\alpha,\beta+1\right)}\left(\xi\right) & =\frac{\lambda-1}{\kappa}Z_{\kappa}^{\left(\alpha,\beta\right)}\left(\xi\right)\nonumber \\
 & -\frac{\Gamma\left(\lambda-\nicefrac{1}{2}\right)}{\kappa^{\beta+2}\Gamma\left(\sigma-\nicefrac{3}{2}\right)}\xi.\label{eq:kPDF-RR_betas}
\end{align}
\end{subequations}

\paragraph{Derivatives.}

Deriving (\ref{eq:Kappa-PDF}) with respect to $\xi$ and then integrating
by parts, one obtains the first derivative\begin{subequations}\label{eq:kPDF-derivatives}
\begin{equation}
Z_{\kappa}^{\left(\alpha,\beta\right)\prime}\left(\xi\right)=-2\left[\frac{\Gamma\left(\lambda-\nicefrac{1}{2}\right)}{\kappa^{\beta+1}\Gamma\left(\sigma-\nicefrac{3}{2}\right)}+\xi Z_{\kappa}^{\left(\alpha,\beta+1\right)}\left(\xi\right)\right].\label{eq:kPDF-1st_derivative}
\end{equation}
The Maxwellian limit of (\ref{eq:kPDF-1st_derivative}) is the well-known
formula\cite{FriedConte61} $Z'\left(\xi\right)=-2\left[1+\xi Z\left(\xi\right)\right]$.

Applying now the operator $d^{n}/d\xi^{n}$ on (\ref{eq:Kappa-PDF})
and integrating by parts, one arrives at a first recurrence relation
for the derivatives, 
\begin{equation}
\begin{aligned}Z_{\kappa}^{\left(\alpha,\beta\right)\left(n+1\right)}\left(\xi\right)= & -2\left[\xi Z_{\kappa}^{\left(\alpha,\beta+1\right)\left(n\right)}\left(\xi\right)\right.\\
 & \left.+nZ_{\kappa}^{\left(\alpha,\beta+1\right)\left(n-1\right)}\left(\xi\right)\right]\\
= & -2\frac{d^{n}}{d\xi^{n}}\left[\xi Z_{\kappa}^{\left(\alpha,\beta+1\right)}\left(\xi\right)\right],\;\left(n\geqslant1\right).
\end{aligned}
\label{eq:kPDF-nth_derivative}
\end{equation}
The Maxwellian limit of (\ref{eq:kPDF-nth_derivative}) is also a
known result.\cite{Robinson89/11}

Expressions (\ref{eq:kPDF-derivatives}a,b) are useful, but they relate
$Z_{\kappa}$ functions with different values of $\beta$\@. In order
to obtain same-$\beta$ expressions for the derivatives, one must
first return to (\ref{eq:kPDF-1st_derivative}) and insert (\ref{eq:kPDF-RR_betas})
to obtain 
\begin{align}
\left(1+\frac{\xi^{2}}{\kappa}\right)Z_{\kappa}^{\left(\alpha,\beta\right)\prime}\left(\xi\right)= & -2\left[\frac{\Gamma\left(\lambda-\nicefrac{1}{2}\right)}{\kappa^{\beta+1}\Gamma\left(\sigma-\nicefrac{3}{2}\right)}\right.\nonumber \\
 & \left.+\frac{\lambda-1}{\kappa}\xi Z_{\kappa}^{\left(\alpha,\beta\right)}\left(\xi\right)\right].\label{eq:kPDF-1st_derivative-2}
\end{align}
Deriving now both sides $n\geqslant1$ times with respect to $\xi$,
using Leibniz formula for the derivative,\cite{Roy+-NIST10} and reorganizing,
there results 
\begin{multline}
\left(1+\frac{\xi^{2}}{\kappa}\right)Z_{\kappa}^{\left(\alpha,\beta\right)\prime\prime}\left(\xi\right)=-2\left[\frac{\lambda-1}{\kappa}Z_{\kappa}^{\left(\alpha,\beta\right)}\left(\xi\right)\right.\\
\left.+\xi Z_{\kappa}^{\left(\alpha,\beta\right)\prime}\left(\xi\right)\vphantom{\frac{\lambda}{\kappa}}\right],\label{eq:kPDF-2nd_derivative}
\end{multline}
\vspace*{-\belowdisplayskip}\vspace*{-\abovedisplayskip} 
\begin{multline}
\left(1+\frac{\xi^{2}}{\kappa}\right)Z_{\kappa}^{\left(\alpha,\beta\right)\left(n+1\right)}\left(\xi\right)\\
=-2\left[\frac{\lambda+n-1}{\kappa}\xi Z_{\kappa}^{\left(\alpha,\beta\right)\left(n\right)}\left(\xi\right)\right.\\
\left.+n\left(\frac{\lambda}{\kappa}+\frac{n-3}{2\kappa}\right)Z_{\kappa}^{\left(\alpha,\beta\right)\left(n-1\right)}\left(\xi\right)\right],\,\left(n\geqslant2\right).\label{eq:kPDF-nth_derivative-2}
\end{multline}
\end{subequations}

A general representation for the $n$-th derivative of $Z_{\kappa}^{\left(\alpha,\beta\right)}\left(\xi\right)$
is also obtainable. Applying the operator $d^{n}/d\xi^{n}$ on the
form (\ref{eq:kPDF_rep2}) and using the formula (\ref{eq:2F1-nth_derivative}),
the following expression if found, 
\begin{multline}
\frac{Z_{\kappa}^{\left(\alpha,\beta\right)\left(n\right)}\left(\xi\right)}{i^{n+1}n!}=\frac{\Gamma\left(\lambda+\nicefrac{n}{2}-1\right)\Gamma\left(\lambda+\nicefrac{n}{2}-\nicefrac{1}{2}\right)}{\kappa^{\beta+\left(n+1\right)/2}\Gamma\left(\sigma-\nicefrac{3}{2}\right)\Gamma\left(\lambda+n\right)}\\
\times\pFq 21\left[{n+1,2\left(\lambda-1\right)+n\atop \lambda+n};\frac{1}{2}\left(1+\frac{i\xi}{\kappa^{1/2}}\right)\right].\label{eq:kPDF-nth_derivative-3}
\end{multline}
Another representation is given by (\ref{eq:kPDF-nth_derivative-4}).

\subsection{The superthermal plasma gyroradius function}

In the same gist that the plasma dispersion function quantifies the
(linear) wave-particle interactions in a finite-temperature plasma,
leading to substantial effects on wave dispersion and absorption,
the \emph{plasma gyroradius function} (PGF), so called in this work,
affects wave propagation at arbitrary angles when the particle's cyclotron
(or Larmor) radius is finite.

In a thermal, Maxwellian plasma, the gyroradius function has a well-known
and simple form, given by Eq. (\ref{eq:Thermal_PGF})\@. In a kappa
plasma, on the other hand, despite the large number of papers already
present in the literature, the corresponding $\kappa$PGF was not
categorized and systematically studied until Ref. \onlinecite{GaelzerZiebell14/12}
proposed a first definition and derived some basic properties. In
this section, a more systematic study of the $\kappa$PGF is made,
thereby complementing the initial formulation given by Ref. \onlinecite{GaelzerZiebell14/12}.

\paragraph{Definition and basic properties.}

The kappa (or superthermal) PGF is defined by the integral 
\begin{equation}
\mathcal{H}_{n,\kappa}^{\left(\alpha,\beta\right)}\left(z\right)=2\int_{0}^{\infty}dx\,\frac{xJ_{n}^{2}\left(yx\right)}{\left(1+x^{2}/\kappa\right)^{\lambda-1}},\:\left(y^{2}=2z\right).\label{eq:Kappa-PGF}
\end{equation}
where the parameters $\alpha$, $\beta$ and $\lambda$ are the same
as for the $\kappa$PDF, whereas $n=\dots,-2,-1,0,1,2,\dots$ is the
harmonic (of the gyrofrequency) number. For plasma physics applications,
the function $\mathcal{H}_{n,\kappa}^{\left(\alpha,\beta\right)}\left(z\right)$
describes the effects of finite particle gyroradius, and thus the
argument will be $z=\mu_{a}$, where $\mu_{a}=k_{\perp}^{2}\rho_{a}^{2}$,
with $\rho_{a}$ being the said Larmor radius.

The definition (\ref{eq:Kappa-PGF}) is slightly different from the
definition given by Ref. \onlinecite{GaelzerZiebell14/12}, but
the significance and importance are the same.

At the Maxwellian limit, if one employs identity (\ref{eq:Exp-limit}),
the remaining integral can be found in any table of mathematical formulæ,
and one readily obtains (\ref{eq:Thermal_PGF})\@. Here, this limit
will be demonstrated by the general representation of the function
$\mathcal{H}_{n,\kappa}^{\left(\alpha,\beta\right)}\left(z\right)$,
which will be given below.

The first property to be derived is the value at the origin $\left(z=0\right)$\@.
At that point, direct integration shows that\cite{OlverMaximon-NIST10}
\begin{equation}
\mathcal{H}_{n,\kappa}^{\left(\alpha,\beta\right)}\left(0\right)=\frac{\kappa}{\lambda-2}\delta_{n,0},\quad\left(\lambda>2\right),\label{eq:kPGF-origin}
\end{equation}
where the definition of the beta function\cite{AskeyRoy-NIST10} was
also employed, and where $\delta_{n,m}$ is the Kronecker delta.

Since $J_{-n}\left(z\right)=\left(-\right)^{n}J_{n}\left(z\right)$
for integer $n$, another straightforward property is $\mathcal{H}_{-n,\kappa}^{\left(\alpha,\beta\right)}\left(z\right)=\mathcal{H}_{n,\kappa}^{\left(\alpha,\beta\right)}\left(z\right)$,
which is shared by the function $\mathscr{H}_{n}\left(z\right)$ as
well.

\paragraph{Representations for $\mathcal{H}_{n,\kappa}^{\left(\alpha,\beta\right)}\left(z\right)$\@.}

When encumbered with the task of finding a computable representation
for $\mathcal{H}_{n,\kappa}^{\left(\alpha,\beta\right)}\left(z\right)$,
one could argue, quite naturally, that since the function depends
on the particle gyroradius, and since in many practical applications
this quantity can be considered small, a ``first-order'' approximation
for $\mathcal{H}_{n,\kappa}^{\left(\alpha,\beta\right)}\left(z\right)$
could be obtained by first expanding $J_{n}^{2}\left(z\right)$ in
a power series, and then evaluating the resulting integrals for the
first few terms. This is an usual procedure to obtain a small-Larmor-radius
approximation for the function $\mathscr{H}_{n}\left(z\right)$.

Applying this method blindly to the $\kappa$PGF, however, one would
invariably obtain a result that either imposes an artificially high
constraint on the value of the $\kappa$ parameter, or is plainly
incorrect, from both the mathematical and physical points of view.
The reason for this unfortunate outcome is simple. Considering an
arbitrary value for the harmonic number $n$, the lowest-order approximation
for the Bessel function is\cite{OlverMaximon-NIST10} 
\[
J_{n}^{2}\left(z\right)\simeq\frac{1}{\left(\left|n\right|!\right)^{2}}\left(\frac{z}{2}\right)^{2\left|n\right|}.
\]
Inserting this approximation into (\ref{eq:Kappa-PGF}), one would
obtain, to lowest order, 
\[
\mathcal{H}_{n,\kappa}^{\left(\alpha,\beta\right)}\left(z\right)\simeq\frac{\kappa\Gamma\left(\lambda-\left|n\right|-2\right)}{2^{\left|n\right|}\left|n\right|!\Gamma\left(\lambda-1\right)}\left(\kappa z\right)^{\left|n\right|},
\]
which only exists for $\lambda>\left|n\right|+2$\@. Higher-order
terms can be included from the series for $J_{n}^{2}\left(z\right)$,
with the outcome that the term of order $k>0$ will be proportional
to $\Gamma\left(\lambda-\left|n\right|-k-2\right)$, thereby imposing
the even stricter constraint $\lambda>\left|n\right|+k+2$.

Consequently, it would appear as if for a plasma where Larmor radius
effects are important up to order $K\geqslant0$ and thermal effects
demand the inclusion of up to $N\geqslant0$ harmonics, the particles
in such a plasma could only have a VDF with $\lambda>N+K+2$, where
the lowest possible value is $\lambda>2$, the constraint already
imposed by Eq. (\ref{eq:kPGF-origin}). Moreover the minimum value
for the parameter $\kappa$, resulting from this constraint, would
be linked with the harmonic number $n$\@. This is an undesirable
and altogether unobserved restriction to the allowable values for
the $\kappa$ index, which has been measured to be as low as $\kappa\simeq2$,
both in the solar wind and in the magnetosphere.\cite{Maksimovic+97/05,Stverak+09/05,Benson+13/08}

On the other hand, recalling that the asymptotic behavior of the Bessel
function for large argument is\cite{OlverMaximon-NIST10} 
\[
J_{n}^{2}\left(z\right)\sim\frac{2}{\pi z}\cos^{2}\left(z-\frac{1}{2}\left|n\right|\pi-\frac{1}{4}\pi\right)\leqslant\frac{2}{\pi z},
\]
the ensuing asymptotic behavior of the integration in (\ref{eq:Kappa-PGF})
is 
\[
\mathcal{H}_{n,\kappa}^{\left(\alpha,\beta\right)}\left(z\right)\sim\frac{\Gamma\left(\lambda-\nicefrac{3}{2}\right)}{\Gamma\left(\lambda-1\right)}\sqrt{\frac{2\kappa}{\pi z}},
\]
showing the constraint $\lambda>\nicefrac{3}{2}$, independent on
the harmonic number and near the observed values of $\kappa$\@.
Hence, the exact value of the integral in (\ref{eq:Kappa-PGF}) must
always exist for any harmonic number $n$, and the correct constraint
imposed over $\kappa$ must be independent of $n$ and close to the
small values observed in the interplanetary environment.

Additionaly, the assumption that the kappa gyroradius function can
be somehow expressed as a simple power series, as is the case with
the function $\mathscr{H}_{n}\left(z\right)$, is simply incorrect.
As it will be shown below, for certain values of the $\lambda$ (or
$\kappa$) parameter, the function $\mathcal{H}_{n,\kappa}^{\left(\alpha,\beta\right)}\left(z\right)$
displays a logarithmic behavior, not present in the Maxwellian PGF\@.
It will also be shown that even when $\mathcal{H}_{n,\kappa}^{\left(\alpha,\beta\right)}\left(z\right)$
is represented by a power series, the lowest-order terms contain a
contribution proportional to a noninteger power. These terms are completely
overlooked by the simplified approach described above. As a consequence,
even if one could argue that the $\kappa$ parameter is small enough
to justify a power-series approximation from the outset, the result
would be incorrect due to the lacking terms.

Therefore, it is the contention in this work, as it was in Ref. \onlinecite{GaelzerZiebell14/12},
that approximate expressions for the kappa gyroradius function can
only be sought after an exact, closed-form representation is found.
It is that representation that will provide the mathematically (and
physically) correct approximation for the function.

Starting again from definition (\ref{eq:Kappa-PGF}), the gyroradius
function $\mathcal{H}_{n,\kappa}^{\left(\alpha,\beta\right)}\left(z\right)$
will now be represented in terms of the Meijer $G$-function discussed
in section \ref{sub:G-function}\@. First, inserting representation
(\ref{eq:G-rep:J_nu^2}) into (\ref{eq:Kappa-PGF}), one can write
\[
\mathcal{H}_{n,\kappa}^{\left(\alpha,\beta\right)}\left(z\right)=\frac{\kappa^{\lambda-1}}{\sqrt{\pi}}\int_{0}^{\infty}dw\,\frac{G_{1,3}^{1,1}\left[2zw\left|{\nicefrac{1}{2}\atop n,-n,0}\right.\right]}{\left(\kappa+w\right)^{\lambda-1}},
\]
after a simple transformation of variable. Identifying this integral
in formula (\ref{eq:G-Lu75_5.6.7}) and using identity (\ref{eq:G-pars_cancel}),
there results 
\begin{equation}
\mathcal{H}_{n,\kappa}^{\left(\alpha,\beta\right)}\left(z\right)=\frac{\pi^{-1/2}\kappa}{\Gamma\left(\lambda-1\right)}G_{1,3}^{2,1}\left[2\kappa z\left|{\nicefrac{1}{2}\atop \lambda-2,n,-n}\right.\right].\label{eq:kPGF-G}
\end{equation}

Formula (\ref{eq:kPGF-G}) is the exact, closed-form representation
of the kappa plasma gyroradius function. According to the definition
of the $G$-function, the only restriction imposed by (\ref{eq:kPGF-G})
is $\lambda\neq\nicefrac{3}{2},\nicefrac{1}{2},-\nicefrac{1}{2},\dots$\@.
Hence, the overall constraint imposed over $\lambda$ is still determined
by (\ref{eq:kPGF-origin}), namely, $\lambda>2$\@. For the ST91
form, this implies that $\kappa>0$ or 1, depending solely on the
parameter $\beta$ and quite within the measured values for $\kappa$
in the solar wind VDFs.

Using formula (\ref{eq:kPGF-G}) and the definition (\ref{eq:Meijer-G}),
one obtains another proof of the Maxwellian limit. Accordingly, 
\begin{gather*}
\begin{aligned}\lim_{\kappa\to\infty}\mathcal{H}_{n,\kappa}^{\left(\alpha,\beta\right)}\left(z\right)= & \frac{1}{\sqrt{\pi}}\frac{1}{2\pi i}\int_{L}ds\frac{\Gamma\left(n-s\right)\Gamma\left(\nicefrac{1}{2}+s\right)}{\Gamma\left(1+n+s\right)}\\
 & \times\lim_{\kappa\to\infty}\left[\frac{\kappa\Gamma\left(\lambda-2-s\right)}{\Gamma\left(\lambda-1\right)}\left(2\kappa z\right)^{s}\right]
\end{aligned}
\\
\hphantom{\mathcal{H}_{n,\kappa}^{\left(\alpha,\beta\right)}\left(z\right)}\begin{aligned} & =\frac{1}{\sqrt{\pi}}\frac{1}{2\pi i}\int_{L}\frac{\Gamma\left(n-s\right)\Gamma\left(\nicefrac{1}{2}+s\right)}{\Gamma\left(1+n+s\right)}\left(2z\right)^{s}ds\\
 & =\frac{1}{\sqrt{\pi}}G_{1,2}^{1,1}\left[2z\left|{\nicefrac{1}{2}\atop n,-n}\right.\right].
\end{aligned}
\end{gather*}
Identifying with (\ref{eq:G-rep:H_nu}) and (\ref{eq:Thermal_PGF}),
one concludes that indeed $\mathcal{H}_{n,\kappa}^{\left(\alpha,\beta\right)}\left(z\right)\xrightarrow{\kappa\to\infty}\mathscr{H}_{n}\left(z\right).$

Looking now for computable representations, the function $\mathcal{H}_{n,\kappa}^{\left(\alpha,\beta\right)}\left(z\right)$
is divided in two cases, depending on the nature of parameter $\lambda$:
integer or noninteger. The second case is handled first. 
\begin{widetext}
If $\lambda-2-n$ is noninteger, then formula (\ref{eq:G-from_pFq})
is valid and one obtains 
\begin{multline}
\mathcal{H}_{n,\kappa}^{\left(\alpha,\beta\right)}\left(z\right)=\frac{\pi^{-1/2}\kappa}{\Gamma\left(\lambda-1\right)}\left[\frac{\Gamma\left(n+2-\lambda\right)\Gamma\left(\lambda-\nicefrac{3}{2}\right)}{\Gamma\left(\lambda-1+n\right)}\left(2\kappa z\right)^{\lambda-2}\pFq 12\left({\lambda-\nicefrac{3}{2}\atop \lambda-1-n,\lambda-1+n};2\kappa z\right)\right.\\
\left.+\frac{\Gamma\left(\lambda-n-2\right)\Gamma\left(n+\nicefrac{1}{2}\right)}{\Gamma\left(2n+1\right)}\left(2\kappa z\right)^{n}\pFq 12\left({n+\nicefrac{1}{2}\atop n+3-\lambda,2n+1};2\kappa z\right)\right],\label{eq:kPGF-1F2}
\end{multline}
where the $\pFq 12\bigl({a\atop b,c};z\bigr)$ function is defined
by (\ref{eq:1F2-def}). 
\end{widetext}

Two important observations about (\ref{eq:kPGF-1F2}) can be made.
First, one can clearly observe that if $\lambda$ is integer, either
term will always contain a singularity. Fortunately, they cancel out
and the result can be written in terms of known functions. This case
will be treated below. The other observation is that when $\lambda$
is not integer, if on writes (\ref{eq:kPGF-1F2}) explicitly as a
power series, as shown in Eq. (\ref{eq:kPGF-power_series}), each
term in the expansion will be proportional to a noninteger power of
$z$.

The second case occurs when $\lambda$ is integer, \emph{i. e.}, $\lambda=2+k$
$\left(k=1,2,\dots\right)$\@. In order to show that now the $\kappa$PGF
is also represented by known functions, the limit $\lambda=2$ will
be temporarily considered in (\ref{eq:kPGF-G}), in which case formula
(\ref{eq:G-rep:IK}) shows that 
\begin{align*}
\mathcal{H}_{n,\kappa}^{\left(\alpha,\beta\right)}\left(z\right) & \stackrel{(\lambda=2)}{=}\frac{\kappa}{\sqrt{\pi}}G_{1,3}^{2,1}\left[2\kappa z\left|{\nicefrac{1}{2}\atop 0,n,-n}\right.\right]\\
 & =2\kappa K_{n}\left(\sqrt{2\kappa z}\right)I_{n}\left(\sqrt{2\kappa z}\right),
\end{align*}
where $K_{n}\left(z\right)$ is the second modified Bessel function\@.\cite{OlverMaximon-NIST10}
This result shows clearly that in this case the function has a logarithmic
singularity since, for $z\approx0$,\cite{OlverMaximon-NIST10} 
\[
K_{n}\left(z\right)\simeq\frac{1}{2}\Gamma\left(n\right)\left(\frac{z}{2}\right)^{-n}+\left(-\right)^{n}\ln\left(\frac{z}{2}\right)I_{n}\left(z\right).
\]

\begin{widetext}
Now, in order to obtain the expression for physical values of $\lambda=3,4,\dots$,
one must take into account the formula (\ref{eq:G-derivatives}),
the Leibniz differentiation formula,\cite{Roy+-NIST10} and the identities\cite{Brychkov08}
\begin{align*}
\frac{\partial^{n}}{\partial z^{n}}\left[z^{\pm\nu/2}I_{\nu}\left(a\sqrt{z}\right)\right] & =\left(\frac{a}{2}\right)^{n}z^{\left(\pm\nu-n\right)/2}I_{\nu\mp n}\left(a\sqrt{z}\right)\\
\frac{\partial^{n}}{\partial z^{n}}\left[z^{\pm\nu/2}K_{\nu}\left(a\sqrt{z}\right)\right] & =\left(-\frac{a}{2}\right)^{n}z^{\left(\pm\nu-n\right)/2}K_{\nu\mp n}\left(a\sqrt{z}\right).
\end{align*}
After some simple algebra, one obtains 
\begin{equation}
\mathcal{H}_{n,\kappa}^{\left(\alpha,\beta\right)}\left(z\right)=\frac{2\kappa}{\Gamma\left(\lambda-1\right)}\left(\frac{\kappa z}{2}\right)^{\lambda/2-1}\sum_{s=0}^{\lambda-2}\left(-\right)^{s}{\lambda-2 \choose s}K_{n-\left(\lambda-2\right)+s}\left(\sqrt{2\kappa z}\right)I_{n+s}\left(\sqrt{2\kappa z}\right),\label{eq:kPGF-IK}
\end{equation}
which is valid for $\lambda=2+k$ $\left(k=1,2,\dots\right)$. 
\end{widetext}

With identities (\ref{eq:kPGF-1F2}) and (\ref{eq:kPGF-IK}), all
possible values for the parameters and the argument are covered and
adequate approximations for $\mathcal{H}_{n,\kappa}^{\left(\alpha,\beta\right)}\left(z\right)$
can be obtained. As an example, in Ref. \onlinecite{GaelzerZiebell14/12}
dispersion relations for dispersive Alfvén waves propagating in a
kappa plasma were derived from suitable approximations to the $\kappa$PGF\@.
From the computational point of view, representations (\ref{eq:kPGF-1F2})
and (\ref{eq:kPGF-IK}) also come in handy. The Meijer $G$-function
is supported by some computer algebra software and also by the \texttt{mpmath}
library.\cite{mpmath} However, for computationally-intensive applications,
codes written in \texttt{Fortran }or \texttt{C}/\texttt{C++} are better
suited.

In order to evaluate the function $\mathcal{H}_{n,\kappa}^{\left(\alpha,\beta\right)}\left(z\right)$,
a special code written in \texttt{Modern Fortran}\cite{Metcalf+11}
was developed. A complete description of the code structure will not
be given here. Suffice it to say that for the case $\lambda$ integer,
given by (\ref{eq:kPGF-IK}), library routines that evaluate the modified
Bessel functions were employed, whereas for noninteger $\lambda$,
although the series (\ref{eq:kPGF-1F2}) formally converges for any
$0<z<\infty$, rounding errors corrupt the accuracy of the result
for $2\kappa z\gtrsim10$ and other strategies are needed. The code
that was written evaluates $\mathcal{H}_{n,\kappa}^{\left(\alpha,\beta\right)}\left(z\right)$
in roughly the same time-frame taken by any library routine that evaluates
a transcendental function, with an accuracy of the order of the machine
precision.

Figure \ref{fig:Hnkab-n0,1} shows some plots of the function $\mathcal{H}_{n,\kappa}^{\left(0,0\right)}\left(z\right)$,
evaluated with the code that was developed. The Maxwellian form $\mathscr{H}_{n}\left(z\right)$
is clearly reached for $\kappa\gg1$ and the departure from the Maxwellian
increases as $\kappa$ diminishes, as expected. The plots also show
that the greatest relative departure occurs for small $z$ $\left(z\lesssim1\right)$,
implying that the effect of the $\kappa$VDF is more pronounced on
small-gyroradius particles.

\begin{figure}
\includegraphics[width=1\columnwidth]{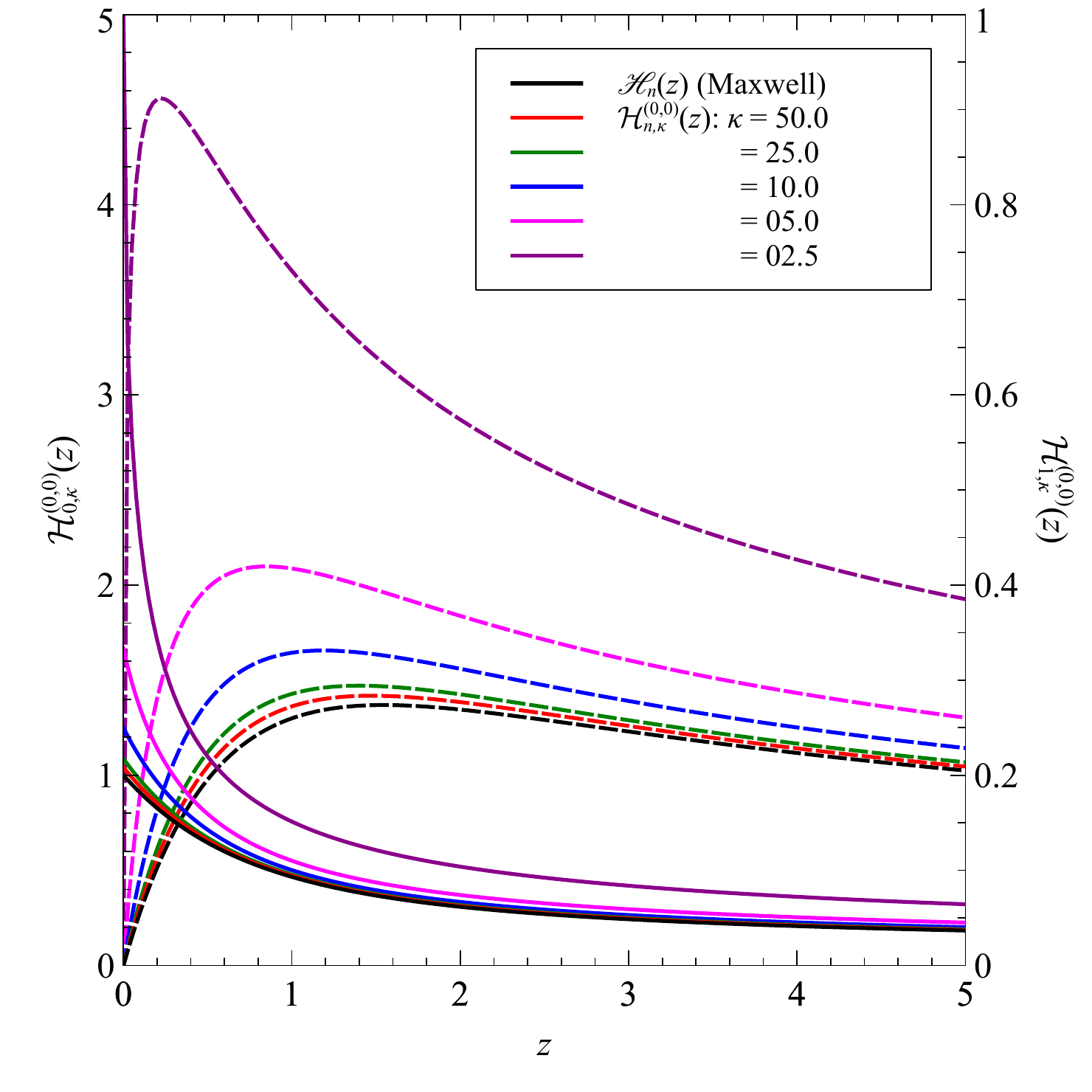}

\caption{Plots of function $\mathcal{H}_{n,\kappa}^{\left(0,0\right)}\left(z\right)$
for $n=0$ (full lines) and $n=1$ (dashed lines) and several values
of the $\kappa$ parameter. The limiting form $\mathscr{H}_{n}\left(z\right)$
for $\kappa\to\infty$ is also included.\label{fig:Hnkab-n0,1}}
\end{figure}

\begin{figure}
\includegraphics[width=1\columnwidth]{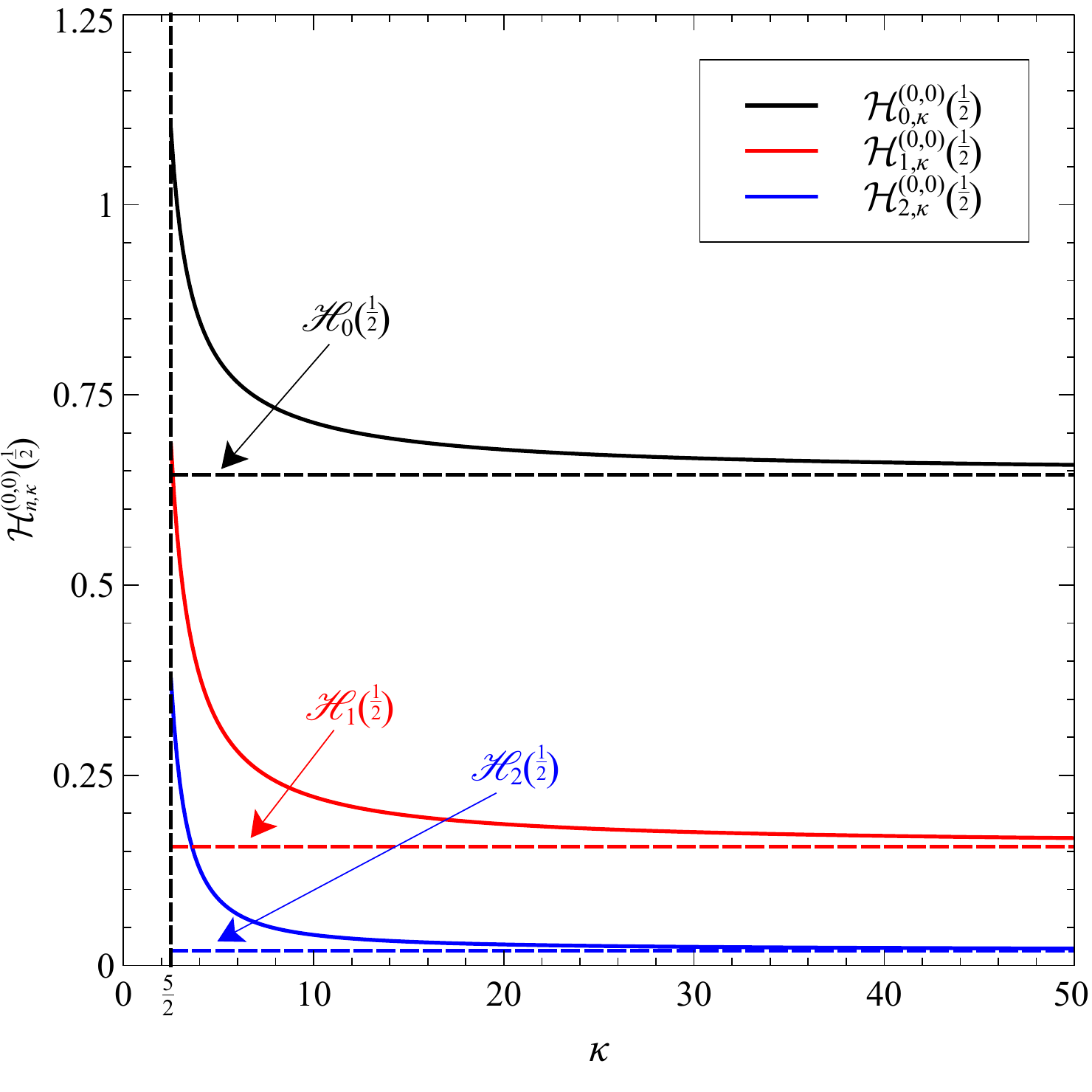}

\caption{Plots of $\mathcal{H}_{n,\kappa}^{\left(0,0\right)}\left(\nicefrac{1}{2}\right)$
as a function of $\kappa$ for $n=0,1,2$\@. The limit value $\mathscr{H}_{n}\left(\nicefrac{1}{2}\right)$
is also shown.\label{fig:Hnkab_vk}}
\end{figure}

A situation where rounding errors could be important in the developed
code might occur when $\lambda$ is in the vicinity of an integer,
when the code still evaluates formula (\ref{eq:kPGF-1F2}), but near
the poles of the gamma functions. However, for the test cases considered,
the code demonstrated a robust performance, as ilustrated by Fig.
\ref{fig:Hnkab_vk}, which shows plots of $\mathcal{H}_{n,\kappa}^{\left(0,0\right)}\left(\nicefrac{1}{2}\right)$
varying $\kappa$, for some values of $n$\@. The graph shows that
the code provides smooth results for all $\nicefrac{5}{2}\leqslant\kappa\leqslant50$.

\paragraph{Derivative and recurrence relations.}

As the last mathematical properties for the $\kappa$PGF, their recurrence
relations will be deducted now.

Starting from the representation (\ref{eq:kPGF-G}), if one writes
down the explicit Mellin-Barnes integral from (\ref{eq:Meijer-G})
and evaluates the derivative of the argument, one obtains 
\begin{multline*}
\frac{d}{dy}G_{1,3}^{2,1}\left[y\left|{\nicefrac{1}{2}\atop \lambda-2,n,-n}\right.\right]\\
=\frac{y^{-1}}{2\pi i}\int_{L}\frac{\Gamma\left(\lambda-2-s\right)\Gamma\left(n-s\right)\Gamma\left(\nicefrac{1}{2}+s\right)}{\Gamma\left(n+1+s\right)}sy^{s}ds.
\end{multline*}
Since $\Gamma\left(n-s\right)=\left(n-s\right)^{-1}\Gamma\left(n+1-s\right)$,
adding and substracting $n$ in the integrand above and then changing
the integration variable as $s\to s+1$, there results 
\begin{multline*}
\frac{d}{dy}G_{1,3}^{2,1}\left[y\left|{\nicefrac{1}{2}\atop \lambda-2,n,-n}\right.\right]\\
=ny^{-1}G_{1,3}^{2,1}\left[y\left|{\nicefrac{1}{2}\atop \lambda-2,n,-n}\right.\right]\\
-\frac{1}{2\pi i}\int_{L}\frac{\Gamma\left(\lambda-3-s\right)\Gamma\left(n-s\right)\Gamma\left(\nicefrac{3}{2}+s\right)}{\Gamma\left(n+2+s\right)}y^{s}ds.
\end{multline*}
Finally, writing $1=\left(n+1+s\right)-\left(n-s\right)$, reorganizing
the terms and identifying the result back to (\ref{eq:kPGF-G}), one
obtains\begin{subequations}\label{eq:kPGF-RR} 
\begin{multline}
\mathcal{H}_{n,\kappa}^{\left(\alpha,\beta\right)\prime}\left(z\right)=\frac{n}{z}\mathcal{H}_{n,\kappa}^{\left(\alpha,\beta\right)}\left(z\right)\\
-\frac{\kappa}{\lambda-2}\left[\mathcal{H}_{n,\kappa}^{\left(\alpha,\beta-1\right)}\left(z\right)-\mathcal{H}_{n+1,\kappa}^{\left(\alpha,\beta-1\right)}\left(z\right)\right],\label{eq:kPGF-RR-1}
\end{multline}
which has the constraint $\lambda>3$ imposed.

Going back now to the first expression for the derivative, employing
the same identity for the gamma function, but now cancelling the denominator,
one obtains in a similar way 
\begin{multline}
\mathcal{H}_{n,\kappa}^{\left(\alpha,\beta\right)\prime}\left(z\right)=-\frac{n}{z}\mathcal{H}_{n,\kappa}^{\left(\alpha,\beta\right)}\left(z\right)\\
-\frac{\kappa}{\lambda-2}\left[\mathcal{H}_{n,\kappa}^{\left(\alpha,\beta-1\right)}\left(z\right)-\mathcal{H}_{n-1,\kappa}^{\left(\alpha,\beta-1\right)}\left(z\right)\right],\label{eq:kPGF-RR-2}
\end{multline}
which has the same condition $\lambda>3$\@.

Finally, by adding and substracting (\ref{eq:kPGF-RR-1}) and (\ref{eq:kPGF-RR-2}),
one obtains the additional relations 
\begin{multline}
\mathcal{H}_{n,\kappa}^{\left(\alpha,\beta\right)\prime}\left(z\right)=-\frac{\kappa}{\lambda-2}\mathcal{H}_{n,\kappa}^{\left(\alpha,\beta-1\right)}\left(z\right)\\
+\frac{1}{2}\frac{\kappa}{\lambda-2}\left[\mathcal{H}_{n-1,\kappa}^{\left(\alpha,\beta-1\right)}\left(z\right)+\mathcal{H}_{n+1,\kappa}^{\left(\alpha,\beta-1\right)}\left(z\right)\right],\label{eq:kPGF-RR-3}
\end{multline}
and 
\begin{multline}
2\frac{n}{z}\mathcal{H}_{n,\kappa}^{\left(\alpha,\beta\right)}\left(z\right)\\
=\frac{\kappa}{\lambda-2}\left[\mathcal{H}_{n-1,\kappa}^{\left(\alpha,\beta-1\right)}\left(z\right)-\mathcal{H}_{n+1,\kappa}^{\left(\alpha,\beta-1\right)}\left(z\right)\right].\label{eq:kPGF-RR-4}
\end{multline}
\end{subequations}

In the Maxwellian limit the recurrence relations (\ref{eq:kPGF-RR}a-d)
will reduce to the corresponding expressions for $\mathscr{H}_{n}\left(z\right)$
and $\mathscr{H}_{n}'\left(z\right)$ that can be obtained from the
recurrence relations of the modified Bessel function.\cite{OlverMaximon-NIST10}

Further properties of the function $\mathcal{H}_{n,\kappa}^{\left(\alpha,\beta\right)}\left(z\right)$
are given in appendix \ref{sec:Additional_props-kappa_plasma_funcs}.

\subsection{The two-variable kappa plasma functions\label{sub:2_vars-PF}}

After discussing at length about the functions $Z_{\kappa}^{\left(\alpha,\beta\right)}\left(\xi\right)$
and $\mathcal{H}_{n,\kappa}^{\left(\alpha,\beta\right)}\left(z\right)$,
one is finally ready to return to the two-variable functions $\mathcal{Z}_{n,\kappa}^{\left(\alpha,\beta\right)}\left(\mu,\xi\right)$
and $\mathcal{Y}_{n,\kappa}^{\left(\alpha,\beta\right)}\left(\mu,\xi\right)$
that appear in the Stix parameters (\ref{eq:Stix_pars_kappa}a-d)
for a kappa plasma.

In order to define these functions, one has to first go back to Eq.
(\ref{eq:DT_Cartesian}), introduce the $\kappa$VDF (\ref{eq:Kappa-VDF})
into the integrals, evaluate the derivatives, identify the rotated
components via Eq. (\ref{eq:DT_rotated}) and then finally identify
the Stix parameters in (\ref{eq:Stix_pars_kappa})\@. By an adequate
change of integration variables and some algebra, one can verify that
the functions in question can be defined as\begin{subequations}\label{eq:ZY_cal-defs}
\begin{align}
\mathcal{Z}_{n,\kappa}^{\left(\alpha,\beta\right)}\left(\mu,\xi\right)= & 2\int_{0}^{\infty}dx\,\frac{xJ_{n}^{2}\left(\nu x\right)}{\left(1+x^{2}/\kappa\right)^{\lambda-1}}\nonumber \\
 & \times Z_{\kappa}^{\left(\alpha,\beta\right)}\left(\frac{\xi}{\sqrt{1+x^{2}/\kappa}}\right)\label{eq:Z_cal}\\
\mathcal{Y}_{n,\kappa}^{\left(\alpha,\beta\right)}\left(\mu,\xi\right)= & \frac{2}{\mu}\int_{0}^{\infty}dx\,\frac{x^{3}J_{n-1}\left(\nu x\right)J_{n+1}\left(\nu x\right)}{\left(1+x^{2}/\kappa\right)^{\lambda-1}}\nonumber \\
 & \times Z_{\kappa}^{\left(\alpha,\beta\right)}\left(\frac{\xi}{\sqrt{1+x^{2}/\kappa}}\right),\label{eq:Y_cal}
\end{align}
\end{subequations}where $\nu^{2}=2\mu$.

Applying the limit $\kappa\to\infty$, one obtains the expressions
\begin{align*}
\mathcal{Z}_{n,\kappa}^{\left(\alpha,\beta\right)}\left(\mu,\xi\right)\to & 2Z_{\kappa}^{\left(\alpha,\beta\right)}\left(\xi\right)\int_{0}^{\infty}dx\,xe^{-x^{2}}J_{n}^{2}\left(\nu x\right)\\
\mathcal{Y}_{n,\kappa}^{\left(\alpha,\beta\right)}\left(\mu,\xi\right)\to & \frac{2}{\mu}Z_{\kappa}^{\left(\alpha,\beta\right)}\left(\xi\right)\\
 & \times\int_{0}^{\infty}dx\,x^{3}e^{-x^{2}}J_{n-1}\left(\nu x\right)J_{n+1}\left(\nu x\right),
\end{align*}
that can be found in any table of integrals. The final result will
be Eqs. (\ref{eq:ZY_cal-Maxwell}).

As it was already mentioned, functions similar to $\mathcal{Z}_{n,\kappa}^{\left(\alpha,\beta\right)}\left(\mu,\xi\right)$
and $\mathcal{Y}_{n,\kappa}^{\left(\alpha,\beta\right)}\left(\mu,\xi\right)$,
also defined in terms of a remaining integral, were already considered
by Ref. \onlinecite{Summers+94/06} and have recently been numerically
implemented by Ref. \onlinecite{Astfalk+15/09}\@. However, the
advantages of having exact, analytically closed-form expressions for
these functions are plenty. First of all, they simplify the derivation
of adequate approximations for dispersion relations, damping or growth-rate
coefficients, among other quantities related to wave propagation.
Moreover, analytical and computable representations are usually advantageous
for numerical applications, both in terms of computing time and accuracy.
As an example, the availability of analytical expressions for the
kappa gyroradious function allowed the authors of Ref. \onlinecite{GaelzerZiebell14/12}
to obtain physically correct expressions for the dispersion relations
of dispersive Alfvén waves.

As it will be shown presently, the two-variable functions defined
above have contributions from both one-variable functions $Z_{\kappa}^{\left(\alpha,\beta\right)}\left(\xi\right)$
and $\mathcal{H}_{n,\kappa}^{\left(\alpha,\beta\right)}\left(z\right)$\@.
It was shown here that the former can always be written in terms of
the Gauss function; hence, it is hypergeometric (\emph{i. e.}, given
by a power series) in nature. However, the second one is not in general
of the same nature. Therefore, the functions $\mathcal{Z}_{n,\kappa}^{\left(\alpha,\beta\right)}\left(\mu,\xi\right)$
and $\mathcal{Y}_{n,\kappa}^{\left(\alpha,\beta\right)}\left(\mu,\xi\right)$
are not representable in general by any two-variable hypergeometric
function, such as the Appell or Horn series\@.\cite{Erdelyi+53a}

The analytical, closed-form representations sought for functions $\mathcal{Z}_{n,\kappa}^{\left(\alpha,\beta\right)}\left(\mu,\xi\right)$
and $\mathcal{Y}_{n,\kappa}^{\left(\alpha,\beta\right)}\left(\mu,\xi\right)$
can be derived in the following manner. Considering first the function
$\mathcal{Z}_{n,\kappa}^{\left(\alpha,\beta\right)}$ in (\ref{eq:Z_cal}),
if the quantity $\xi$ is a point inside the domain of the principal
branch of $Z_{\kappa}^{\left(\alpha,\beta\right)}\left(\zeta\right)$,
then, as the integration is carried out, the argument $\zeta\left(x\right)=\xi/\sqrt{1+x^{2}/\kappa}$
follows a curve in the complex plane attaching the point $\zeta\left(0\right)=\xi$
with the origin. Since it is always possible to find a region $R$
where the function $Z_{\kappa}^{\left(\alpha,\beta\right)}\left(\zeta\right)$
is analytic and that contains the whole integration path, Taylor's
theorem assures that it is a possible to expand the function in a
power series around any interior point of $R$. Therefore, for any
$0\leqslant x<\infty$, the function $Z_{\kappa}^{\left(\alpha,\beta\right)}\left(\zeta\left(x\right)\right)$
can be expanded in a power series around $\zeta=\xi$, as 
\begin{multline}
Z_{\kappa}^{\left(\alpha,\beta\right)}\left(\frac{\xi}{\sqrt{1+x^{2}/\kappa}}\right)=\sum_{k=0}^{\infty}\sum_{\ell=0}^{k}\left(-\right)^{\ell}{k \choose \ell}\frac{\left(-\xi\right)^{k}}{k!}\\
\times Z_{\kappa}^{\left(\alpha,\beta\right)\left(k\right)}\left(\xi\right)\left(1+\frac{x^{2}}{\kappa}\right)^{-\ell/2}.\label{eq:Z_k^(a,b)-Z_cal-series}
\end{multline}

Inserting (\ref{eq:Z_k^(a,b)-Z_cal-series}) into (\ref{eq:Z_cal}),
one can immediately identify the $\mathcal{H}_{n,\kappa}^{\left(\alpha,\beta\right)}$
function from the definition (\ref{eq:Kappa-PGF}) and obtain\begin{subequations}\label{eq:ZY_cal-ZH}
\begin{multline}
\mathcal{Z}_{n,\kappa}^{\left(\alpha,\beta\right)}\left(\mu,\xi\right)=\sum_{k=0}^{\infty}\sum_{\ell=0}^{k}\left(-\right)^{\ell}{k \choose \ell}\frac{\left(-\xi\right)^{k}}{k!}\\
\times Z_{\kappa}^{\left(\alpha,\beta\right)\left(k\right)}\left(\xi\right)\mathcal{H}_{n,\kappa}^{\left(\alpha,\beta+\ell/2\right)}\left(\mu\right).\label{eq:Z_cal-ZH}
\end{multline}

Following the same procedure for $\mathcal{Y}_{n,\kappa}^{\left(\alpha,\beta\right)}\left(\mu,\xi\right)$
in (\ref{eq:Y_cal}), one obtains 
\begin{multline}
\mathcal{Y}_{n,\kappa}^{\left(\alpha,\beta\right)}\left(\mu,\xi\right)=\sum_{k=0}^{\infty}\sum_{\ell=0}^{k}{k \choose \ell}\frac{\left(-\right)^{\ell}\kappa}{\lambda+\nicefrac{\ell}{2}-2}\frac{\left(-\xi\right)^{k}}{k!}\\
\times Z_{\kappa}^{\left(\alpha,\beta\right)\left(k\right)}\left(\xi\right)\mathcal{H}_{n,\kappa}^{\left(\alpha,\beta+\ell/2-1\right)\prime}\left(\mu\right),\label{eq:Y_cal-ZH}
\end{multline}
\end{subequations}where the identity (\ref{eq:Hnkab-derivative-identity})
was also employed.

Although expressions (\ref{eq:ZY_cal-ZH}a,b) look formidable, they
allow a significant speed-up for numerical evaluations of functions
$\mathcal{Z}_{n,\kappa}^{\left(\alpha,\beta\right)}\left(\mu,\xi\right)$
and $\mathcal{Y}_{n,\kappa}^{\left(\alpha,\beta\right)}\left(\mu,\xi\right)$\@.
A short test was performed, comparing the evaluation of $\mathcal{Z}_{n,\kappa}^{\left(\alpha,\beta\right)}\left(\mu,\xi\right)$
both via numerical quadrature applied to formula (\ref{eq:Z_cal})
and by truncating the $k$-series in (\ref{eq:Z_cal-ZH}) to a value
$k_{\mathrm{max}}\geqslant0$.

\begin{figure}
\includegraphics[width=1\columnwidth]{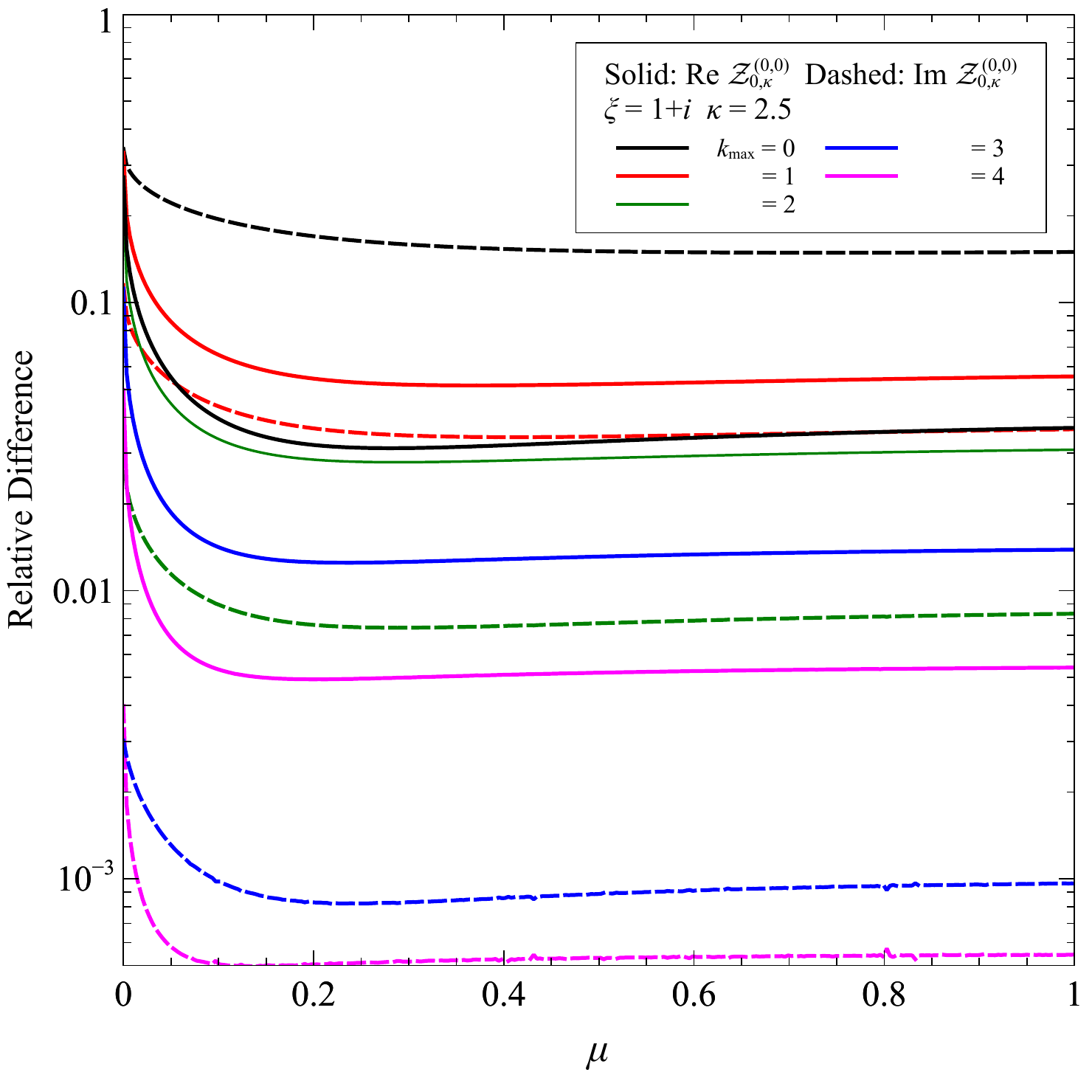}

\caption{Plots of the real (solid lines) and imaginary (dashed) parts of the
relative difference $\Delta$ between functions $\mathcal{Z}_{\mathrm{sum}}$
and $\mathcal{Z}_{\mathrm{int}}$ for $n=0$, $\alpha=\beta=0$, $\kappa=2.5$,
$\xi=1+i$ and different values of $k_{\mathrm{max}}$.\label{fig:Z_cal-RelDiff-n0}}
\end{figure}

Call $\mathcal{Z}_{\mathrm{sum}}\left(\mu\right)$ the result obtained
by truncating the series (\ref{eq:Z_cal-ZH}) up to $k=k_{\mathrm{max}}$
and $\mathcal{Z}_{\mathrm{int}}\left(\mu\right)$ the result of (\ref{eq:Z_cal}),
evaluated via an adaptive quadrature routine, with requested relative
accuracy of $\epsilon=10^{-5}$\@. This will be considered the ``correct''
value, against which the truncated series result can be compared.
figure \ref{fig:Z_cal-RelDiff-n0} shows plots of the relative difference
$\Delta=\left|1-\mathcal{Z}_{\mathrm{sum}}/\mathcal{Z}_{\mathrm{int}}\right|$,
both for the real and imaginary parts of the functions. The results
were obtained by keeping fixed $n=0$, $\alpha=\beta=0$, $\kappa=2.5$
and $\xi=1+i$, and varying $\mu$\@.

\begin{figure}
\includegraphics[width=1\columnwidth]{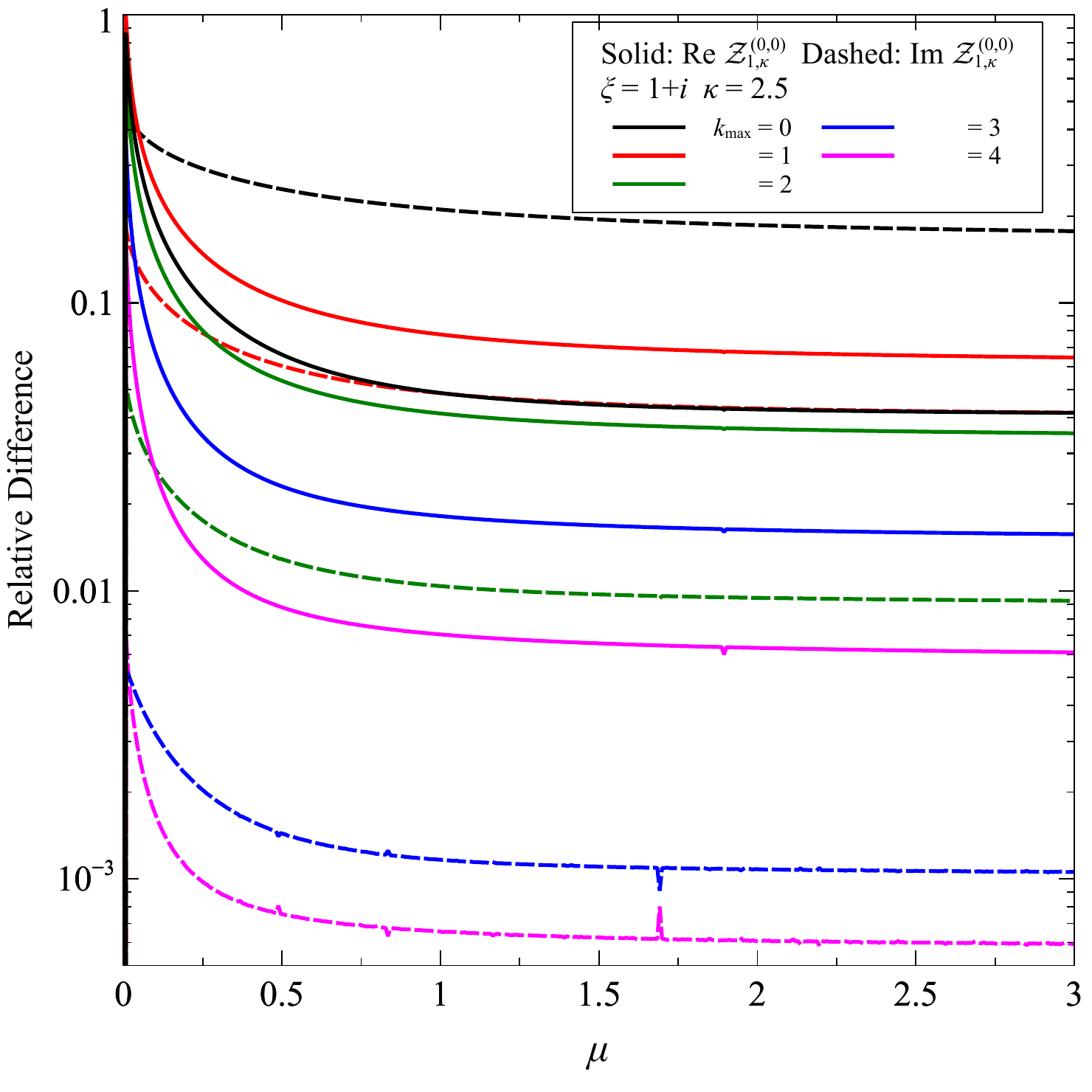}

\caption{Same as Fig. \ref{fig:Z_cal-RelDiff-n0}, but for $n=1$.\label{fig:Z_cal-RelDiff-n1}}
\end{figure}

Figure \ref{fig:Z_cal-RelDiff-n1} shows the same results, but now
for the harmonic number $n=1$\@. The results show that indeed the
relative difference decreases with $k_{\mathrm{max}}$\@. Interestingly,
the approximation with $k_{\mathrm{max}}=0$ was indeed better than
with $k_{\mathrm{max}}=1$ in both figures, but from $k_{\mathrm{max}}=2$
the quantity $\Delta$ steadily decreases, for both the real and imaginary
parts, by roughly one order of magnitude for each successive value
of $k_{\mathrm{max}}$\@.

In both figures, for $k_{\mathrm{max}}=4$ one gets roughly $\Delta_{r}\simeq10^{-3}$
and $\Delta_{i}\simeq10^{-4}$\@. If one were to plot both $\mathcal{Z}_{\mathrm{int}}\left(\mu\right)$
and $\mathcal{Z}_{\mathrm{sum}}\left(\mu\right)$ for $k_{\mathrm{max}}=4$
in the same slide, the graph of the latter would be almost indistinguishable
from the former; however, calling $T_{\mathrm{sum}}$ and $T_{\mathrm{int}}$
the average computing time per point for $\mathcal{Z}_{\mathrm{sum}}$
and $\mathcal{Z}_{\mathrm{int}}$, respectively, for $n=1$ and $k_{\mathrm{max}}=4$
in Fig. \ref{fig:Z_cal-RelDiff-n1}, the ratio of time per point resulted
$T_{\mathrm{sum}}/T_{\mathrm{int}}\simeq10^{-2}$\@. In (\ref{eq:Z_cal-ZH}),
the total number of $Z_{\kappa}^{\left(\alpha,\beta\right)\left(k\right)}\left(\xi\right)$
evaluations is $k_{\mathrm{max}}+1$, whereas the total number of
$\mathcal{H}_{n,\kappa}^{\left(\alpha,\beta+\ell/2\right)}\left(\mu\right)$
evaluations is $\frac{1}{2}\left(k_{\mathrm{max}}+1\right)\left(k_{\mathrm{max}}+2\right)$
(15, for $k_{\mathrm{max}}=4$)\@. Nevertheless, the evaluation of
$\mathcal{Z}_{\mathrm{sum}}\left(\mu\right)$ was in average almost
100 times faster than the evaluation $\mathcal{Z}_{\mathrm{int}}\left(\mu\right)$\@.
The speed-up can be further increased if one employs the recurrence
relations for $Z_{\kappa}^{\left(\alpha,\beta\right)\left(k\right)}\left(\xi\right)$
and $\mathcal{H}_{n,\kappa}^{\left(\alpha,\beta+\ell/2\right)}\left(\mu\right)$,
derived in the previous sections, which were not used in the tests.
However, in this case a stability analysis of the recurrence relations
must be first carried out.

Therefore, although Eqs. (\ref{eq:ZY_cal-defs}a,b) are simpler to
implement, for computer-intensive applications one should employ the
closed-forms shown by (\ref{eq:ZY_cal-ZH}a,b)\@. As was also argued,
these last representations also allow for the correct derivation of
approximations, valid for some particular range of particle species,
frequencies, propagation angles and wave polarization.

\section{Numerical applications\label{sec:Numerical-applications}}

As a simple application of the formalism developed in the preceding
sections, the dispersion equation (\ref{eq:DE_rotated}) was solved
for some particular cases.

Since the intention was to solely provide a simple demonstration of
the formalism, it was assumed an electron-proton plasma, both formally
described by the same distribution function (\ref{eq:Kappa-VDF}),
and both in the ST91 form. However, the dispersion equation was solved
only for high-frequency waves propagating at arbitrary angles with
$\boldsymbol{B}_{0}$\@. Hence, the protons only serve to provide
a stationary background by taking $m_{p}\to\infty$ ($m_{p}$: proton
mass)\@. The Stix parameters in Eqs. (\ref{eq:Stix_pars_kappa})
were evaluated with $a=e$ only and by truncating the harmonic number
series to $-n_{\mathrm{max}}\leqslant n\leqslant n_{\mathrm{max}}$\@.
In all solutions presented below, $n_{\mathrm{max}}=1$.

The kappa plasma functions $\mathcal{Z}_{n,\kappa}^{\left(\alpha,\beta\right)}\left(\mu,\xi\right)$
and $\mathcal{Y}_{n,\kappa}^{\left(\alpha,\beta\right)}\left(\mu,\xi\right)$
and their derivatives in Eqs. (\ref{eq:Stix_pars_kappa}a-d) were
evaluated by (\ref{eq:ZY_cal-ZH}a,b) on the lowest possible order,
\emph{i. e.}, $k_{\mathrm{max}}=0$\@. In this case, the exact expressions
(\ref{eq:ZY_cal-ZH}a,b) are replaced by the approximations 
\begin{align*}
\widetilde{\mathcal{Z}}_{n,\kappa}^{\left(1,\beta\right)}\left(\mu,\xi\right) & =\mathcal{H}_{n,\kappa}^{\left(1,\beta\right)}\left(\mu\right)Z_{\kappa}^{\left(1,\beta\right)}\left(\xi\right)\\
\widetilde{\mathcal{Y}}_{n,\kappa}^{\left(1,\beta\right)}\left(\mu,\xi\right) & =\frac{\kappa}{\lambda-2}\mathcal{H}_{n,\kappa}^{\left(1,\beta-1\right)\prime}\left(\mu\right)Z_{\kappa}^{\left(1,\beta\right)}\left(\xi\right),
\end{align*}
in which case their formal structure is the same as the Maxwellian
limits (\ref{eq:ZY_cal-Maxwell})\@. One must point out here that
although $\widetilde{\mathcal{Z}}_{n,\kappa}^{\left(1,\beta\right)}\left(\mu,\xi\right)$
and $\widetilde{\mathcal{Y}}_{n,\kappa}^{\left(1,\beta\right)}\left(\mu,\xi\right)$
are approximations when the distribution function is the isotropic
$\kappa$VDF given by (\ref{eq:Kappa-VDF}), or when one employs a
bi-kappa model, if one employs instead another anisotropic distribution
model, such as the product-bi-kappa or kappa-Maxwellian VDFs, which
describe statistical distributions of particles with uncorrelated
velocity directions, the same structure will correspond to the \emph{exact}
expressions for the functions $\mathcal{Z}_{n,\kappa}^{\left(\alpha,\beta\right)}\left(\mu,\xi\right)$
and $\mathcal{Y}_{n,\kappa}^{\left(\alpha,\beta\right)}\left(\mu,\xi\right)$,
since the expansion (\ref{eq:Z_k^(a,b)-Z_cal-series}) won't be needed
in these cases.

\begin{figure}
\begin{minipage}[t]{1\columnwidth}%

\begin{center}
\includegraphics[width=1\columnwidth]{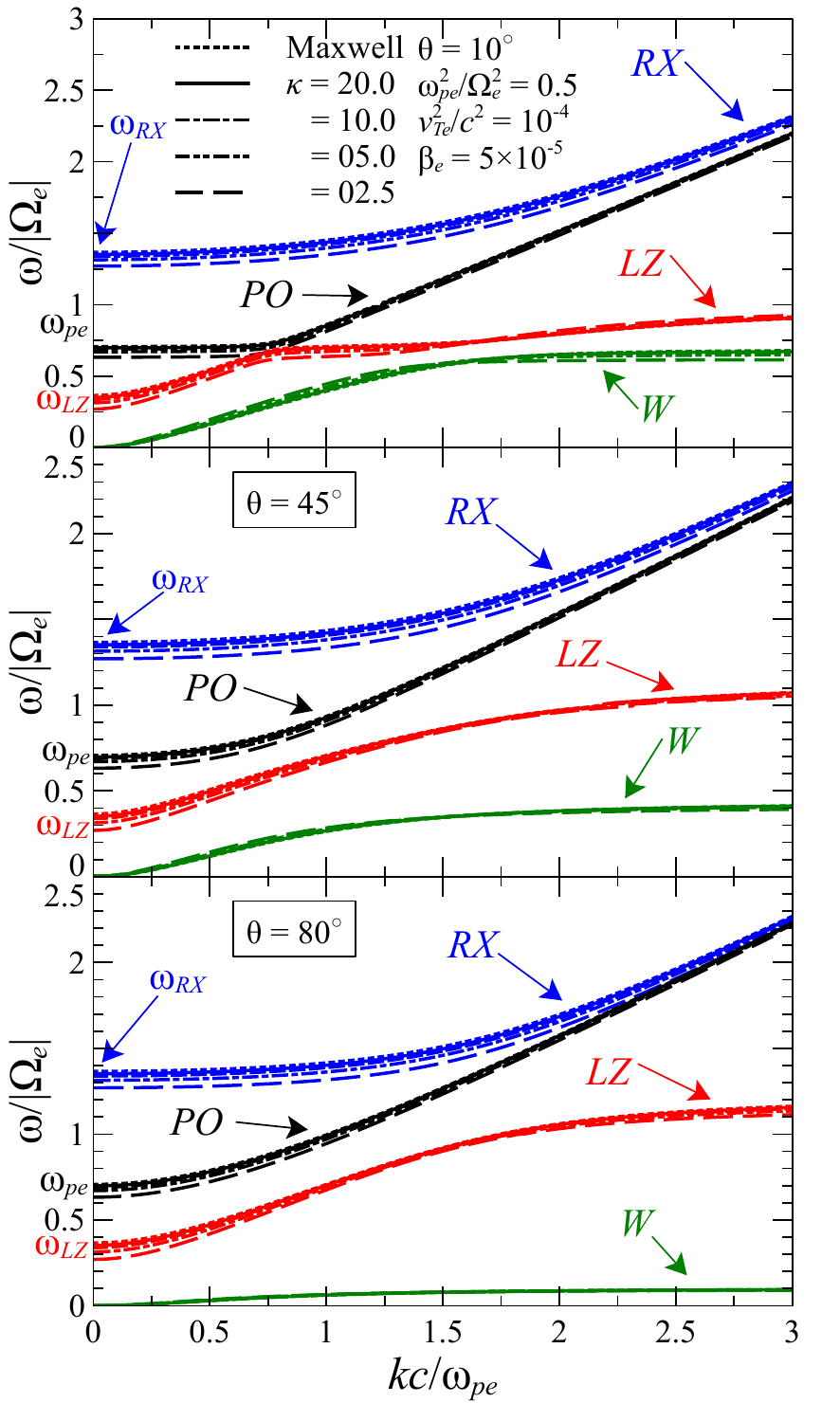} 
\par\end{center}

\caption{Plots of high-frequency dispersion relations in a low-beta, kappa
plasma, as the numerical solutions of the dispersion equation (\ref{eq:DE_rotated}),
for propagation angles $\theta=10^{\circ}$, $45^{\circ}$ and $80^{\circ}$\@.
For each angle, the dispersion equation is solved for several values
of the $\kappa$ parameter, including the limit $\kappa\to\infty$
(Maxwell)\@. The modes with elliptic polarization are named: $RX$
(right-fast extraordinary), $PO$ (plasma-ordinary), $LZ$ (left-slow
extraordinary) and $W$ (whistler).\label{fig:obk_hf_sb}}
\end{minipage}
\end{figure}

Figure \ref{fig:obk_hf_sb} shows the numerical solutions of the dispersion
equation (\ref{eq:DE_rotated}), \emph{i. e.}, the dispersion relations,
for high-frequency waves propagating at three oblique propagation
angles, relative to $\boldsymbol{B}_{0}$: $\theta=10^{\circ}$, $45^{\circ}$
and $80^{\circ}$\@. For each angle, the dispersion equation is solved
for different values of the $\kappa$ parameter, including the Maxwellian
limit. The other physical parameters adopted in Fig. \ref{fig:obk_hf_sb}
are: $\omega_{pe}^{2}/\Omega_{e}^{2}=0.5$, corresponding to a low-density
plasma, and $v_{Te}^{2}/c^{2}=10^{-4}$\@. Both these quantities
determine the electron beta parameter 
\[
\beta_{e}=\frac{\omega_{pe}^{2}}{\Omega_{e}^{2}}\frac{v_{Te}^{2}}{c^{2}}=\frac{n_{e}T_{e}}{B_{0}^{2}/8\pi}=5\times10^{-5},
\]
which measures the ratio of the thermal to magnetic field energy densities.
Hence, the dispersion relations in Fig. \ref{fig:obk_hf_sb} are typical
to a low-beta plasma.

Since for oblique propagation the polarizations of the eigenmodes
are elliptic, in this work we adopted a simplistic nomenclature that
combines the names for both exactly parallel or perpendicular directions.
In the slide for $\theta=10^{\circ}$, for instance, the blue curves
are labelled ``$RX$'' because in the parallel direction case $\left(\theta=0^{\circ}\right)$,
these would be the modes with right-hand circular polarization $\left(R\right)$\@.
On the other hand, at the perpendicular direction $\left(\theta=90^{\circ}\right)$,
they would correspond to the fast extraordinary mode $\left(X\right)$\@.
Hence, this elliptic mode is termed the $RX$, or ``right-fast extraordinary''
mode. The ``$PO$'' mode is the ``plasma-ordinary'' mode, since
for $\theta=0^{\circ}$ this mode would be the longitudinal (plasma,
$P$) mode, whereas for $\theta=90^{\circ}$ it would be the ordinary
$\left(O\right)$ mode. The mode ``$LZ$'' or ``left-slow extraordinary''
would be the left-hand circularly polarized mode $\left(L\right)$
for parallel or the slow extraordinary $\left(Z\right)$ for perpendicular
directions. Finally, the ``$W$'' or ``whistler'' mode is the
lower-frequency branch of the $R$ mode for $\theta=0^{\circ}$, which
disappears as $\theta\to90^{\circ}$, since only the electron inertia
is included in the dispersion equation.

In each panel of Fig. \ref{fig:obk_hf_sb}, the cut-off frequencies
$\omega_{RX}$ and $\omega_{LZ}$ corresponding to the $R$ and $L$
modes cut-offs for a Maxwellian plasma,\cite{Brambilla98} which and
are independent of the propagation angle $\theta$, are also identified.

Since in a low-beta plasma the thermal effects are less important,
the effect of the long tail of the $\kappa$VDF is not very pronounced.
Notwithstanding, the dispersion relations noticeably depart from the
Maxwellian case as $\kappa$ decreases, with the $\kappa=20$ case
practically identical to the Maxwellian limit. One also notices that
the departure is roughly constant in wavenumber for almost all modes,
although the $W$ mode is the least affected by the shape of the VDF\@.

\begin{figure}
\begin{centering}
\includegraphics[width=1\columnwidth]{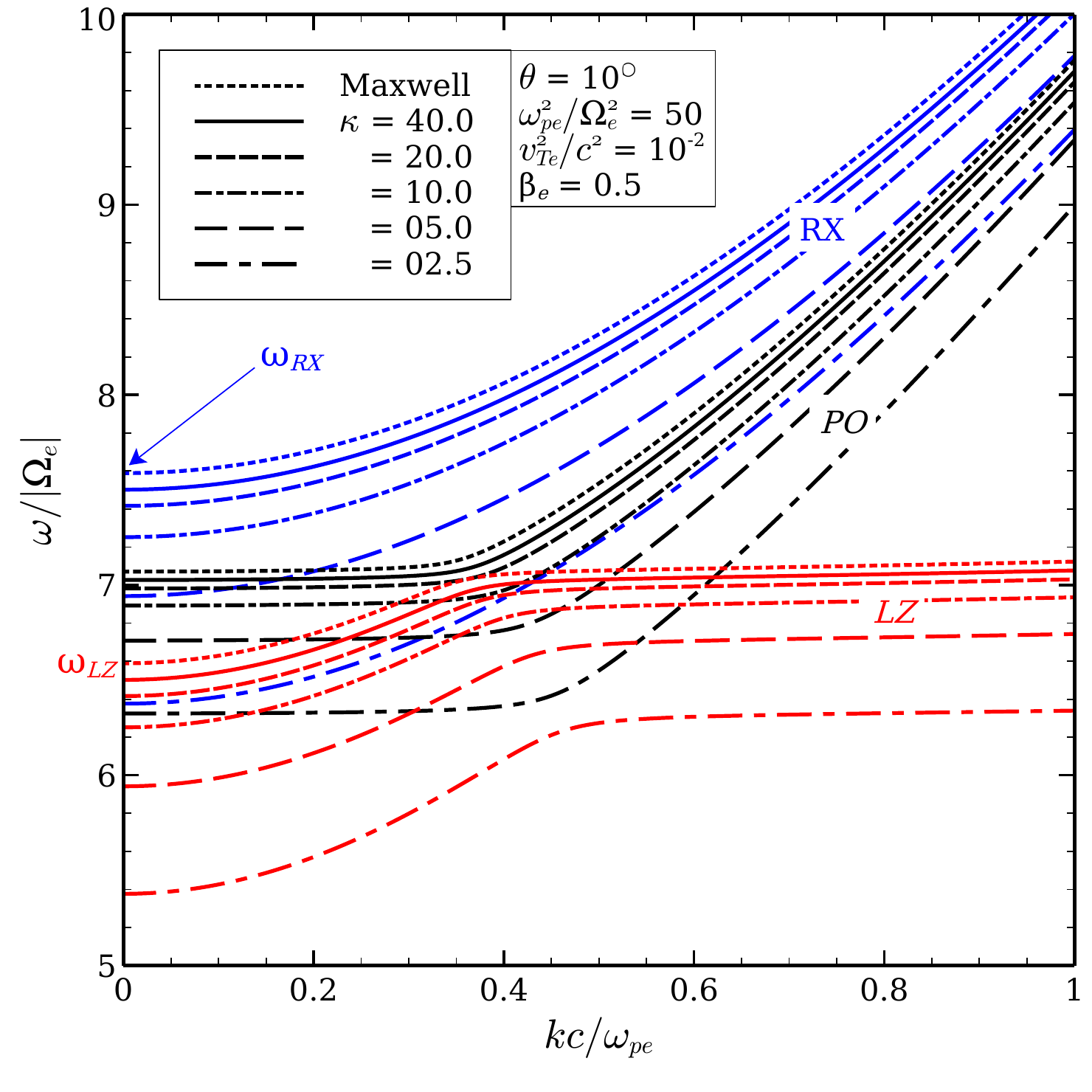}
\par\end{centering}

\caption{Plots of high-frequency dispersion relations in a high-beta, kappa
plasma, as the numerical solutions of the dispersion equation (\ref{eq:DE_rotated}),
for propagation angle $\theta=10^{\circ}$\@. The dispersion equation
is solved for several values of the $\kappa$ parameter, including
the limit $\kappa\to\infty$ (Maxwell)\@. The modes with elliptic
polarization are named: $RX$ (right-fast extraordinary), $PO$ (plasma-ordinary)
and $LZ$ (left-slow extraordinary).\label{fig:obk_hf_hb-t10}}
\end{figure}

\begin{figure}
\begin{centering}
\includegraphics[width=1\columnwidth]{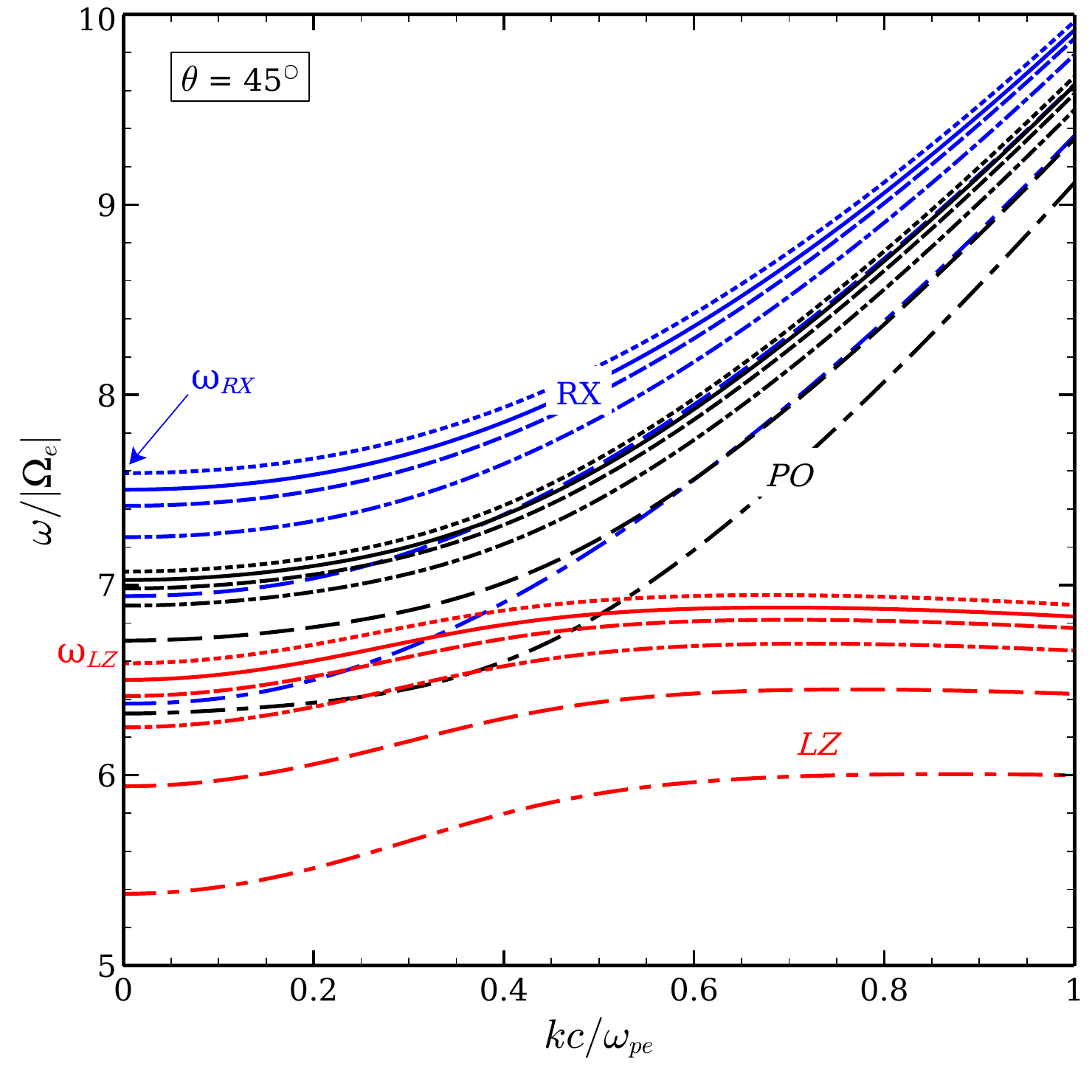}
\par\end{centering}

\caption{Same as Fig. \ref{fig:obk_hf_hb-t10}, but with propagation angle
$\theta=45^{\circ}$.\label{fig:obk_hf_hb-t45}}
\end{figure}

\begin{figure}
\begin{centering}
\includegraphics[width=1\columnwidth]{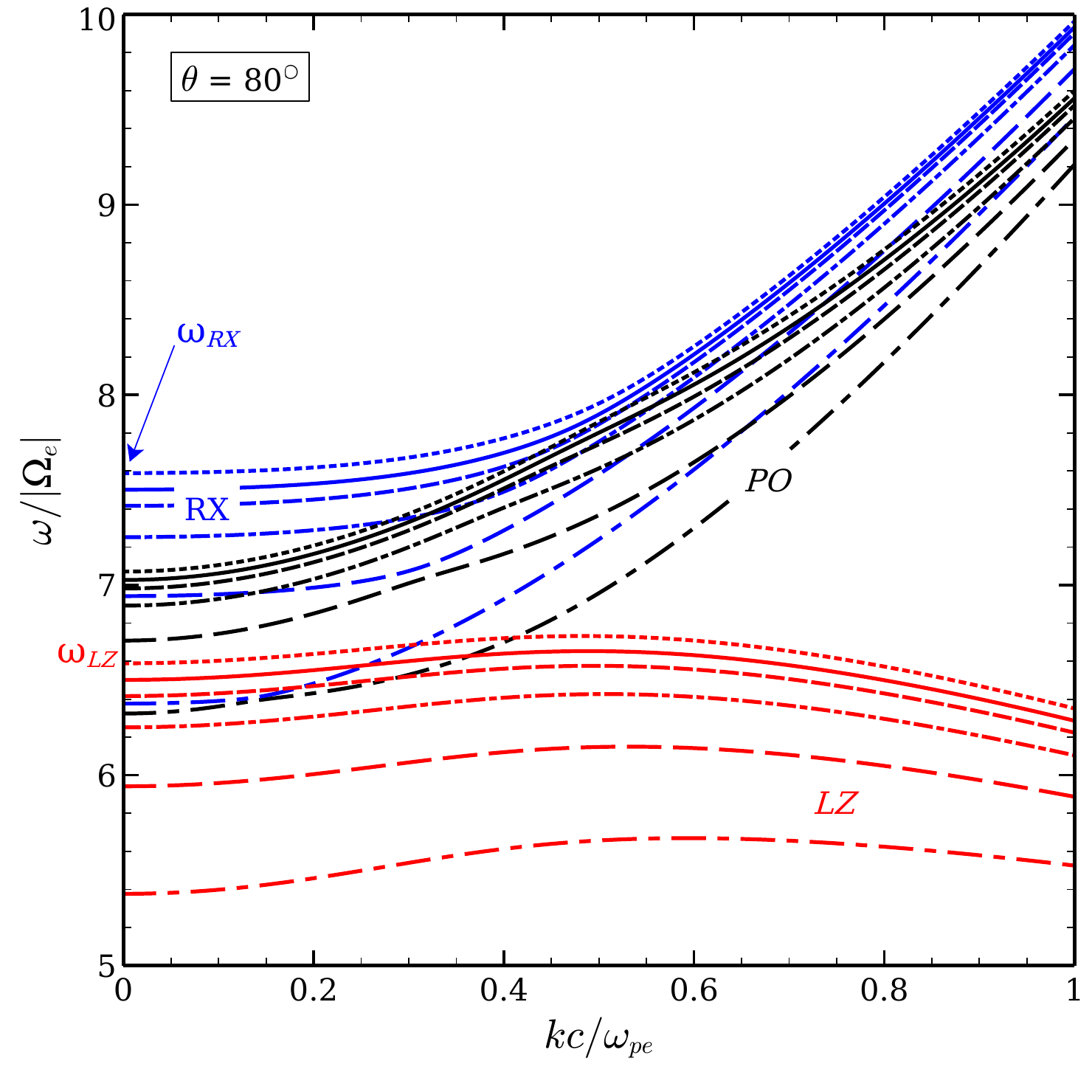}
\par\end{centering}

\caption{Same as Figs. \ref{fig:obk_hf_hb-t10} and \ref{fig:obk_hf_hb-t45},
but with propagation angle $\theta=80^{\circ}$.\label{fig:obk_hf_hb-t80}}
\end{figure}

A quite different scenario appears for a high-density, high-temperature
and high-beta plasma. Figures \ref{fig:obk_hf_hb-t10} - \ref{fig:obk_hf_hb-t80}
show the solutions of (\ref{eq:DE_rotated}) for $\omega_{pe}^{2}/\Omega_{e}^{2}=50$,
$v_{Te}^{2}/c^{2}=10^{-2}$ and, consequently, $\beta_{e}=0.5$\@.
In this case the effect of the superthermal particles on wave dispersion
is considerably more important than it is in a low-beta plasma. Now
the $W$ mode is not shown because it occupies a region well below
the displayed range of frequencies. In all propagation angles, for
a sufficiently small $\kappa$ the dispersion relation for a given
mode and at the same wavenumber can assume a range of frequencies
previously taken by another mode for a larger $\kappa$\@. Therefore,
it is expected that the effect of the superthermal tails on the VDFs
will be significant for moderate to high-beta plasmas in all propagation
angles.

Further studies of oblique waves propagating in superthermal plasmas
will be conducted in future papers, including the effects of anisotropies
in temperature.

\section{Conclusions\label{sec:Conclusions}}

In this work, a general treatment for the problem of wave propagation
in a magnetized superthermal plasma was proposed. The formulation
is valid for any number of particle species, wave frequency and propagation
angle.

The dielectric tensor components are written in terms of thermal Stix
parameters, which in turn are written in terms of special functions
that appear when the velocity distribution functions are isotropic
kappa distributions. The mathematical properties of the special functions
are discussed in detail and several useful identities and formulae
are derived.

As a demonstration of the usefulness of the formulation proposed here,
the dispersion relations for high-frequency waves propagating at several
angles relative to the ambient magnetic field are obtained as the
numerical solutions of the dispersion equation for a kappa plasma.

It is expected that this formulation will be very useful for the study
of wave propagation and amplification/damping in superthermal plasmas,
such as the solar wind or planetary magnetospheres. 
\begin{acknowledgments}
The authors acknowledge support provided by Conselho Nacional de Desenvolvimento
Científico e Tecnológico (CNPq), grants No. 304461/2012-1, 478728/2012-3
and 304363/2014-6. 
\end{acknowledgments}

\appendix
%dummy comment inserted by tex2lyx to ensure that this paragraph is not empty

\section{Additional properties of the kappa plasma functions\label{sec:Additional_props-kappa_plasma_funcs}}

\subsection{Properties of $\boldsymbol{Z_{\kappa}^{\left(\alpha,\beta\right)}\left(\xi\right)}$}

\paragraph{Other representations.}

Employing formula (\ref{eq:2F1_analytic_cont}) and the function (\ref{eq:1F0}),
result (\ref{eq:kPDF_rep1}) can be analytically continued across
the line $\xi_{i}=0$, resulting in the representation 
\begin{multline}
Z_{\kappa}^{\left(\alpha,\beta\right)}\left(\xi\right)=Z_{\kappa,NC}^{\left(\alpha,\beta\right)}\left(\xi\right)\\
+\gamma\frac{2i\sqrt{\pi}\Gamma\left(\lambda-1\right)}{\kappa^{\beta+1/2}\Gamma\left(\sigma-\nicefrac{3}{2}\right)}\left(1+\frac{\xi^{2}}{\kappa}\right)^{-\left(\lambda-1\right)},\label{eq:kPDF_rep1c}
\end{multline}
where $Z_{\kappa,NC}^{\left(\alpha,\beta\right)}\left(\xi\right)$
is the non-continued representation (\ref{eq:kPDF_rep1}), and 
\[
\gamma=\begin{cases}
0, & 0<\arg\xi\leqslant\pi\\
1, & \pi<\arg\xi\leqslant2\pi\text{ (or \ensuremath{-\pi<\arg\xi\leqslant0})}.
\end{cases}
\]

Notice that with (\ref{eq:kPDF_rep1c}), the branch cut moved to the
line $-\sqrt{\kappa}\geqslant\xi_{i}>-\infty$, as is the case of
all other computable representations of $Z_{\kappa}^{\left(\alpha,\beta\right)}\left(\xi\right)$.

It is easy to show that $Z_{\kappa}^{\left(\alpha,\beta\right)}\left(\xi\right)$
always reduces to a polynomial when $\lambda$ is integer. First,
starting from (\ref{eq:kPDF_rep3}) and inserting into (\ref{eq:2F1-transf_linear-a}),
there results 
\begin{multline*}
Z_{\kappa}^{\left(\alpha,\beta\right)}\left(\xi\right)=\left[\frac{i\sqrt{\pi}\Gamma\left(\lambda-1\right)}{\kappa^{\beta+1/2}\Gamma\left(\sigma-\nicefrac{3}{2}\right)}-\frac{2\Gamma\left(\lambda-\nicefrac{1}{2}\right)\xi}{\kappa^{\beta+1}\Gamma\left(\sigma-\nicefrac{3}{2}\right)}\right.\\
\left.\times\pFq 21\left({2-\lambda,\nicefrac{1}{2}\atop \nicefrac{3}{2}};-\frac{\xi^{2}}{\kappa}\right)\right]\left(1+\frac{\xi^{2}}{\kappa}\right)^{-\left(\lambda-1\right)}.
\end{multline*}
According to (\ref{eq:2F1-transf-polynomial-1}), if $\lambda=2+m$
$\left(m=0,1,2,\dots\right)$, the Gauss function reduces to a polynomial
of degree $\lambda-2$\@. This result is also interesting because
its Maxwellian limit is another well-known representation of the Fried
\& Conte function,\cite{FriedConte61} $Z\left(\xi\right)=i\sqrt{\pi}e^{-\xi^{2}}\mathrm{erfc}\left(-i\xi\right)$,
where $\mathrm{erfc}\left(z\right)$ is the complementary error function.\cite{Temme-NIST10-1}

\paragraph{Series representations.}

Several expansions were obtained for the $\kappa$PDF.

\paragraph{Power series.}

Inserting definition (\ref{eq:2F1_series}) into the form (\ref{eq:kPDF_rep3}),
one readily obtains 
\begin{gather}
Z_{\kappa}^{\left(\alpha,\beta\right)}\left(\xi\right)=-\frac{\pi^{1/2}\kappa^{-\beta-1}}{\Gamma\left(\sigma-\nicefrac{3}{2}\right)}\xi\sum_{k=0}^{\infty}\frac{\Gamma\left(\lambda+k-\nicefrac{1}{2}\right)}{\Gamma\left(k+\nicefrac{3}{2}\right)}\left(-\frac{\xi^{2}}{\kappa}\right)^{k}\nonumber \\
+i\frac{\pi^{1/2}\Gamma\left(\lambda-1\right)}{\kappa^{\beta+1/2}\Gamma\left(\sigma-\nicefrac{3}{2}\right)}\left(1+\frac{\xi^{2}}{\kappa}\right)^{-\left(\lambda-1\right)},\label{eq:kPDF-power_series}
\end{gather}
which converges within the radius $\left|\xi^{2}\right|<\kappa$\@.
The Maxwellian limit of (\ref{eq:kPDF-power_series}) reduces, via
convolution, to the well-known series for $Z$$\left(\xi\right)$.\cite{FriedConte61}

\paragraph{Asymptotic series.}

Inserting definition (\ref{eq:2F1_series}) into the form (\ref{eq:kPDF_rep4}),
one readily obtains 
\begin{gather*}
Z_{\kappa}^{\left(\alpha,\beta\right)}\left(\xi\right)=-\frac{\Gamma\left(\lambda-\nicefrac{3}{2}\right)}{\kappa^{\beta}\Gamma\left(\sigma-\nicefrac{3}{2}\right)}\frac{1}{\xi}\sum_{k=0}^{\infty}\frac{\left(\nicefrac{1}{2}\right)_{k}}{\left(\nicefrac{5}{2}-\lambda\right)_{k}}\left(-\frac{\kappa}{\xi^{2}}\right)^{k}\\
+\frac{\pi^{1/2}\Gamma\left(\lambda-1\right)}{\kappa^{\beta+1/2}\Gamma\left(\sigma-\nicefrac{3}{2}\right)}\left[i-\tan\left(\lambda\pi\right)\right]\left(1+\frac{\xi^{2}}{\kappa}\right)^{-\left(\lambda-1\right)}.
\end{gather*}

\begin{widetext}
When $\lambda=\nicefrac{3}{2}+m$ $\left(m=1,2,\dots\right)$, the
result above is not valid, since singularities appear in both terms.
In this case, one must return to (\ref{eq:kPDF_rep3}) and employ
the special case (\ref{eq:2F1-transf-special}), in which case there
results 
\begin{multline*}
Z_{\kappa}^{\left(\alpha,\beta\right)}\left(\xi\right)=-\frac{\pi^{-1/2}\kappa^{-\beta}}{\Gamma\left(\kappa+\alpha-\nicefrac{3}{2}\right)}\frac{1}{\xi}\left\{ \sum_{k=0}^{m-1}\Gamma\left(k+\frac{1}{2}\right)\left(m-k-1\right)!\left(\frac{\kappa}{\xi^{2}}\right)^{k}\right.\\
\left.+\left(\frac{\kappa}{\xi^{2}}\right)^{m}\sum_{k=0}^{\infty}\frac{\Gamma\left(k+m+\nicefrac{1}{2}\right)}{k!}\left[\ln\left(\frac{\xi^{2}}{\kappa}\right)+t_{m,k}\right]\left(-\frac{\kappa}{\xi^{2}}\right)^{k}\right\} \\
+i\pi^{1/2}\frac{\Gamma\left(m+\nicefrac{1}{2}\right)}{\kappa^{\beta+1/2}\Gamma\left(\kappa+\alpha-\nicefrac{3}{2}\right)}\left(1+\frac{\xi^{2}}{\kappa}\right)^{-\left(m+1/2\right)},
\end{multline*}
where 
\[
t_{m,k}=\psi\left(k+1\right)-\psi\left(k+m+\frac{1}{2}\right).
\]
This result shows that there is a logarithmic singularity for the
case $\lambda=m+\nicefrac{3}{2}$ when $\left|\xi^{2}/\kappa\right|\to\infty$. 
\end{widetext}

\paragraph{Derivatives.}

Starting from (\ref{eq:kPDF-nth_derivative-3}) and aplying in sequence
transformations (\ref{eq:2F1-transf_quad(2)}), (\ref{eq:2F1-transf_linear-a})
and (\ref{eq:2F1-transf_linear-c}), one obtains 
\begin{gather}
\frac{Z_{\kappa}^{\left(\alpha,\beta\right)\left(n\right)}\left(\xi\right)}{-i^{n}\sqrt{\pi}n!}=\frac{\kappa^{-\left(\beta+1/2+n/2\right)}}{\Gamma\left(\sigma-\nicefrac{3}{2}\right)}\left[\frac{\Gamma\left(\lambda-\frac{1}{2}+\frac{n}{2}\right)}{\Gamma\left(\frac{1}{2}+\frac{n}{2}\right)}\frac{2\xi}{\sqrt{\kappa}}\right.\nonumber \\
\times\pFq 21\left({\lambda-\frac{1}{2}+\frac{n}{2},1+\frac{n}{2}\atop \nicefrac{3}{2}};-\frac{\xi^{2}}{\kappa}\right)-i\frac{\Gamma\left(\lambda-1+\frac{n}{2}\right)}{\Gamma\left(1+\frac{n}{2}\right)}\nonumber \\
\left.\times\pFq 21\left({\lambda-1+\frac{n}{2},\frac{1}{2}+\frac{n}{2}\atop \nicefrac{1}{2}};-\frac{\xi^{2}}{\kappa}\right)\right],\label{eq:kPDF-nth_derivative-4}
\end{gather}
which immediately renders the power series expansion for $Z_{\kappa}^{\left(\alpha,\beta\right)\left(n\right)}\left(\xi\right)$.

Taking the limit $x\to\infty$ in (\ref{eq:Z_k^(a,b)-Z_cal-series}),
and using the value (\ref{eq:kPDF-origin}), the resulting expression,
\[
\sum_{k=0}^{\infty}\frac{\left(-\xi\right)^{k}}{k!}Z_{\kappa}^{\left(\alpha,\beta\right)\left(k\right)}\left(\xi\right)=i\frac{\sqrt{\pi}\Gamma\left(\lambda-1\right)}{\kappa^{\beta+1/2}\Gamma\left(\sigma-\nicefrac{3}{2}\right)},
\]
is a sum rule for the derivatives of the $\kappa$PDF\@. Once again,
the Maxwellian limit returns a known result,\cite{Robinson89/11}
\[
\sum_{k=0}^{\infty}\frac{\left(-\xi\right)^{k}}{k!}Z^{\left(k\right)}\left(\xi\right)=i\sqrt{\pi}.
\]

\subsection{Properties of $\boldsymbol{\mathcal{H}_{n,\kappa}^{\left(\alpha,\beta\right)}\left(z\right)}$}

An explicit power series expansion when $\lambda$ is not integer
can be written from (\ref{eq:kPGF-1F2}) and (\ref{eq:1F2-def}) as
\begin{equation}
\mathcal{H}_{n,\kappa}^{\left(\alpha,\beta\right)}\left(z\right)=\frac{\kappa\left(2\kappa z\right)^{n}}{\sqrt{\pi}\Gamma\left(\lambda-1\right)}\sum_{k=0}^{\infty}H_{n,k}^{\left(\lambda\right)}\left(z\right)\frac{\left(2\kappa z\right)^{k}}{k!},\label{eq:kPGF-power_series}
\end{equation}
where 
\begin{multline*}
H_{n,k}^{\left(\lambda\right)}\left(z\right)=\frac{\Gamma\left(\lambda-n-2\right)\Gamma\left(n+\nicefrac{1}{2}+k\right)}{\Gamma\left(2n+1+k\right)\left(n+3-\lambda\right)_{k}}\\
+\frac{\Gamma\left(n+2-\lambda\right)\Gamma\left(\lambda-\nicefrac{3}{2}+k\right)}{\Gamma\left(\lambda-1+n+k\right)\left(\lambda-1-n\right)_{k}}\left(2\kappa z\right)^{\lambda-n-2}.
\end{multline*}
Notice that each term in this series is proportional to a noninteger
power of $z$.

Another identity related to the derivative of $\mathcal{H}_{n,\kappa}^{\left(\alpha,\beta\right)}\left(z\right)$
will be obtained now, regarding the integral in the definition (\ref{eq:Y_cal})\@.
Starting from (\ref{eq:G-rep:J_mu-J_nu}) and employing formula (\ref{eq:G-derivatives}),
one obtains 
\begin{multline*}
\int_{0}^{\infty}dx\,\frac{x^{3}J_{n-1}\left(\nu x\right)J_{n+1}\left(\nu x\right)}{\left(1+x^{2}/\kappa\right)^{\lambda-1}}\\
=\frac{\mu}{\sqrt{\pi}}\frac{d}{d\mu}\int_{0}^{\infty}dx\,\frac{x^{3}G_{1,3}^{1,1}\left[2\mu x^{2}\left|{\nicefrac{1}{2}\atop n,-n,-1}\right.\right]}{\left(1+x^{2}/\kappa\right)^{\lambda-1}},
\end{multline*}
recalling that $\nu^{2}=2\mu$\@. Then, using formula (\ref{eq:G-Lu75_5.6.7})
and identifying the result with the definition (\ref{eq:Kappa-PGF}),
one concludes that 
\begin{multline}
\int_{0}^{\infty}dx\,\frac{x^{3}J_{n-1}\left(\nu x\right)J_{n+1}\left(\nu x\right)}{\left(1+x^{2}/\kappa\right)^{\lambda-1}}\\
=\frac{1}{2}\frac{\kappa\mu}{\lambda-2}\mathcal{H}_{n,\kappa}^{\left(\alpha,\beta-2\right)\prime}\left(\mu\right).\label{eq:Hnkab-derivative-identity}
\end{multline}

\section{Hypergeometric functions employed in kappa plasmas\label{sec:Hypergeometric-functions}}

There are two classes of special functions that are employed in this
and other theoretical works concerning superthermal plasmas, namely,
the generalized hypergeometric series and the Meijer $G$ functions.
Some of their properties will be presented here.

\subsection{The generalized hypergeometric series}

The general expression for the hypergeometric series is 
\begin{equation}
\pFq pq\left({a_{1},\dots,a_{p}\atop b_{1},\dots,b_{q}};z\right)=\sum_{k=0}^{\infty}\frac{\left(a_{1}\right)_{k}\cdots\left(a_{p}\right)_{k}}{\left(b_{1}\right)_{k}\cdots\left(b_{q}\right)_{k}}\frac{z^{k}}{k!},\label{eq:General_hypergeom_series}
\end{equation}
where $p,q$ are natural numbers, the sets $\left\{ a_{p}\right\} $,
$\left\{ b_{q}\right\} $ and the argument $z$ are in general complex,
and $\left(\alpha\right)_{n}=\Gamma\left(\alpha+n\right)/\Gamma\left(\alpha\right)$
is the Pochhammer symbol. Unless explicitly stated, all properties
presented here can be found in Ref. \onlinecite{AskeyDaalhuis-NIST10}.

Except when any of the inferior parameters $b_{1},\dots,b_{q}$ is
a nonpositive integer, the hypergeometric series $\pFq pq\left(\cdots;z\right)$
belongs to the class $\mathbb{C}^{p+q+1}$ within its convergence
radius, which divides it in three classes: \emph{(i)} $q\leqslant p$,
\emph{(ii)} $p=q+1$, and \emph{(iii)} $p\geqslant q+2$\@. In this
work, we employ functions of the first two classes. Series of class
3 are not convergent except at the origin. The Meijer $G$-function,
discussed in section \ref{sub:G-function}, lends an analytical representation
for these functions.

\paragraph{Class 1 series: the Kummer and $\protect\pFq 12$ hypergeometric
functions.}

\paragraph{The Kummer confluent hypergeometric function.}

The Kummer function is defined from (\ref{eq:General_hypergeom_series})
as 
\begin{equation}
\pFq 11\left({a\atop b};z\right)=\sum_{k=0}^{\infty}\frac{\left(a\right)_{k}}{\left(b\right)_{k}}\frac{z^{k}}{k!}.\label{eq:Kummer-1F1-series}
\end{equation}

\paragraph{The $\protect\pFq 12\bigl({a\atop b,c};z\bigr)$ hypergeometric function.}

From (\ref{eq:General_hypergeom_series}), 
\begin{equation}
\pFq 12\left({a\atop b,c};z\right)=\sum_{k=0}^{\infty}\frac{\left(a\right)_{k}}{\left(b\right)_{k}\left(c\right)_{k}}\frac{z^{k}}{k!}.\label{eq:1F2-def}
\end{equation}
Notice that both these series, for $\pFq 11$ and $\pFq 12$, converge
for any $\left|z\right|<\infty$\@. Thus, the $\pFq 12\left(\cdots;z\right)$
function belongs to the class $\mathbb{C}^{4}$, is meromorphic on
the $b,c$ planes except at the nonpositive integer points and entire
on the $a,z$ planes.

\paragraph{Class 2 series: the Gauss function.}

The Gauss hypergeometric series is defined from (\ref{eq:General_hypergeom_series})
as\cite{AskeyDaalhuis-NIST10,Daalhuis-NIST10b} 
\begin{equation}
\pFq 21\left({a,b\atop c};z\right)=\sum_{k=0}^{\infty}\frac{\left(a\right)_{k}\left(b\right)_{k}}{\left(c\right)_{k}}\frac{z^{k}}{k!}.\label{eq:2F1_series}
\end{equation}
This series, and all other functions belonging to the same class,
is convergent within the unit circle $\left|z\right|<1$ and conditionally
convergent along it. Some properties of the Gauss function are presented
below.

\paragraph{Convergence at $\left|z\right|=1$ and analyticity.}

Over the unit circle, the series (\ref{eq:2F1_series}) converges:
\emph{(i)} absolutely if $\Re\left(c-a-b\right)>0$, \emph{(ii)} conditionally
if $-1<\Re\left(c-a-b\right)\leqslant0$, and \emph{(iii)} diverges
if $\Re\left(c-a-b\right)\leqslant-1$\@. For $\left|z\right|>1$
the function must be analytically continued.

The Gauss function has a branch point at $z=1$, with the branch line
running over $1\leqslant z<\infty$\@. The principal branch is defined
as the region $0<\arg\left(z-1\right)\leqslant2\pi$.

When $a=-m$ $\left(m=0,1,2,\dots\right)$ and $c\neq0,-1,-2,\dots,$
the function $\pFq 21\bigl({-m,b\atop c};z\bigr)$ reduces to a polynomial
of degree $m$. Obviously, $\pFq 21\bigl(\begin{smallmatrix}a,b\\
c
\end{smallmatrix};z\bigr)=\pFq 21\bigl(\begin{smallmatrix}b,a\\
c
\end{smallmatrix};z\bigr)$.

\paragraph{Integral representation.}

The Gauss function can be expressed, when $\Re c>\Re b>0$, by the
integral 
\begin{multline}
\pFq 21\left({a,b\atop c};z\right)=\frac{\Gamma\left(c\right)}{\Gamma\left(b\right)\Gamma\left(c-b\right)}\\
\times\int_{0}^{1}t^{b-1}\left(1-t\right)^{c-b-1}\left(1-tz\right)^{-a}dt.\label{eq:2F1_integral_rep}
\end{multline}

\paragraph{Analytic continuation.}

Writing $z=x+iy$, the difference of the values of $\pFq 21$ across
the branch line $x>1$ is 
\begin{multline}
\pFq 21\left({a,b\atop c};x+i0\right)-\pFq 21\left({a,b\atop c};x-i0\right)\\
=\frac{2\pi i\Gamma\left(c\right)\left(x-1\right)^{c-a-b}}{\Gamma\left(a\right)\Gamma\left(b\right)\Gamma\left(c-a-b+1\right)}\\
\times\pFq 21\left({c-a,c-b\atop c-a-b+1};1-x\right).\label{eq:2F1_analytic_cont}
\end{multline}

\paragraph{Linear and quadratic transformations.}
\begin{widetext}
The analytical continuation of the series (\ref{eq:2F1_series}) can
also be accomplished employing several known transformations of $\pFq 21\bigl(\begin{smallmatrix}a,b\\
c
\end{smallmatrix};z\bigr)$, either linear or nonlinear\@. Some of these transformations are
given below\begin{subequations}

\begin{align}
\pFq 21\left({a,b\atop c};z\right) & =\left(1-z\right)^{c-a-b}\pFq 21\left({c-a,c-b\atop c};z\right)\label{eq:2F1-transf_linear-a}\\
 & =\frac{\Gamma\left(c\right)\Gamma\left(c-a-b\right)}{\Gamma\left(c-a\right)\Gamma\left(c-b\right)}\pFq 21\left({a,b\atop a+b-c+1};1-z\right)\nonumber \\
 & +\left(1-z\right)^{c-a-b}\frac{\Gamma\left(c\right)\Gamma\left(a+b-c\right)}{\Gamma\left(a\right)\Gamma\left(b\right)}\pFq 21\left({c-a,c-b\atop c-a-b+1};1-z\right)\:\left(\left|\arg\left(1-z\right)\right|<\pi\right)\label{eq:2F1-transf_linear-c}\\
 & =\frac{\Gamma\left(c\right)\Gamma\left(b-a\right)}{\Gamma\left(b\right)\Gamma\left(c-a\right)}\left(-z\right)^{-a}\pFq 21\left({a,1-c+a\atop 1-b+a};\frac{1}{z}\right)\nonumber \\
 & +\frac{\Gamma\left(c\right)\Gamma\left(a-b\right)}{\Gamma\left(a\right)\Gamma\left(c-b\right)}\left(-z\right)^{-b}\pFq 21\left({b,1-c+b\atop 1-a+b};\frac{1}{z}\right)\;\left(\left|\arg\left(-z\right)\right|<\pi\right)\label{eq:2F1-transf_linear-d}\\
\pFq 21\left({a,b\atop a+b-\nicefrac{1}{2}};z\right) & =-\left(1-z\right)^{-1/2}\pFq 21\left({2a-1,2b-1\atop a+b-\nicefrac{1}{2}};\frac{1}{2}+\frac{1}{2}\sqrt{1-z}\right)\label{eq:2F1-transf_quadratic}\\
\pFq 21\left({a,b\atop \frac{1}{2}a+\frac{1}{2}b+\frac{1}{2}};z\right) & =\pFq 21\left({\frac{1}{2}a,\frac{1}{2}b\atop \frac{1}{2}a+\frac{1}{2}b+\frac{1}{2}};4z-4z^{2}\right).\label{eq:2F1-transf_quad(2)}
\end{align}

Some special cases of the transformations above are also relevant
for this work. If $b-a=m$ $\left(m=0,1,2,\dots\right)$, then 
\begin{multline}
\pFq 21\left({a,a+m\atop c};z\right)=\frac{\Gamma\left(c\right)}{\Gamma\left(a+m\right)}\left(-z\right)^{-a}\sum_{k=0}^{m-1}\frac{\left(a\right)_{k}\left(m-k-1\right)!}{k!\Gamma\left(c-a-k\right)}z^{-k}\\
+\frac{\Gamma\left(c\right)}{\Gamma\left(a\right)}\left(-z\right)^{-a}\sum_{k=0}^{\infty}\frac{\left(-\right)^{k}\left(a+m\right)_{k}z^{-k-m}}{k!\left(k+m\right)!\Gamma\left(c-a-k-m\right)}\\
\times\left[\ln\left(-z\right)+\psi\left(k+1\right)+\psi\left(k+m+1\right)-\psi\left(a+k+m\right)-\psi\left(c-a-k-m\right)\right],\;\left({\left|z\right|>1\atop \left|\arg\left(-z\right)\right|<\pi}\right),\label{eq:2F1-transf-special}
\end{multline}
where $\psi\left(z\right)$ is the digamma function.\cite{AskeyRoy-NIST10}

On the other hand, if $a=-m$ $\left(m=0,1,2,\dots\right)$ and neither
$b$ nor $c$ are nonpositive integers, the Gauss function reduces
to the polynomial, 
\begin{equation}
F\left({-m,b\atop c};z\right)=\sum_{n=0}^{m}\frac{\left(-m\right)_{n}\left(b\right)_{n}}{\left(c\right)_{n}}\frac{z^{n}}{n!}=\sum_{n=0}^{m}\binom{m}{n}\frac{\left(b\right)_{n}}{\left(c\right)_{n}}\left(-z\right)^{n}\label{eq:2F1-transf-polynomial-1}
\end{equation}
\end{subequations}
\end{widetext}

\paragraph{Derivatives.}

The general formula for the $n$-th derivative of the Gauss function
is 
\begin{equation}
\frac{d^{n}}{dz^{n}}\pFq 21\left({a,b\atop c};z\right)=\frac{\left(a\right)_{n}\left(b\right)_{n}}{\left(c\right)_{n}}\pFq 21\left({a+n,b+n\atop c+n};z\right).\label{eq:2F1-nth_derivative}
\end{equation}

\paragraph{Other functions in the same class.}

We have also employed the function 
\begin{equation}
\pFq 10\left({a\atop -};z\right)\equiv\pFq 21\left({a,c\atop c};z\right)=\left(1-z\right)^{-a}.\label{eq:1F0}
\end{equation}

\subsection{The Meijer $\boldsymbol{G}$-function\label{sub:G-function}}

The Meijer $G$-function is that function whose Mellin transform\cite{ParisKaminski01}
can be expressed as a ratio of certain products of gamma functions.
Consequently its definition is given by the Mellin-Barnes contour
integral 
\begin{equation}
G_{p,q}^{m,n}\left[z\left|{\left(a_{p}\right)\atop \left(b_{q}\right)}\right.\right]=\frac{1}{2\pi i}\int_{L}\Phi\left({\left(a_{p}\right)\atop \left(b_{q}\right)};s\right)z^{s}ds,\label{eq:Meijer-G}
\end{equation}
where 
\begin{multline*}
\Phi\left({\left(a_{p}\right)\atop \left(b_{q}\right)};s\right)\\
=\frac{\prod_{j=1}^{m}\Gamma\left(b_{j}-s\right)\prod_{j=1}^{n}\Gamma\left(1-a_{j}+s\right)}{\prod_{j=m+1}^{q}\Gamma\left(1-b_{j}+s\right)\prod_{j=n+1}^{p}\Gamma\left(a_{j}-s\right)}.
\end{multline*}
In (\ref{eq:Meijer-G}), $p,q=0,1,2,\dots$, $0\leqslant m\leqslant q$
and $0\leqslant n\leqslant p$\@. If $m+1>q$ or $n+1>p$, the product
is replaced by one. The notation is such that $\left(a_{p}\right)\doteq a_{1},a_{2},\dots,a_{p}$
and $\left(b_{q}\right)\doteq b_{1},b_{2},\dots,b_{q}$\@. It is
assumed that the $a_{j}$'s and $b_{j}$'s are such that no pole of
$\Gamma\left(b_{j}-s\right)$ $\left(j=1,\dots,m\right)$ coincides
with any pole of $\Gamma\left(1-a_{k}+s\right)$ $\left(k=1,\dots,n\right)$,
\emph{i. e.}, $a_{k}-b_{j}\neq1,2,\dots$\@. It is also assumed that
$z\neq0$, since the origin is a branch point.

The integration contour $L$ in (\ref{eq:Meijer-G}) corresponds to
that of the inverse Mellin transform, but deformed in such a way that
the poles of $\Gamma\left(b_{j}-s\right)$ $\left(j=1,\dots,m\right)$
lie to the right of the contour, whereas the poles of $\Gamma\left(1-a_{j}+s\right)$
$\left(j=1,\dots,n\right)$ lie to the left of the same path. A detailed
account on all possible integration paths can be found in Refs. \onlinecite{Luke75,Prudnikov90v3,AskeyDaalhuis-NIST10}\@.
All properties of the $G$-function shown here are likewise found
in these sources.

The $G$-function has remarkable properties. For instance, it contains
all generalized hypergeometric functions, but it also represents functions
that can not be expanded in a power series anywhere, such as functions
with logarithmic singularities. The definition (\ref{eq:Meijer-G})
forms a class that is closed under reflections of the argument ($z\to-z$
and $z\to z^{-1}$), multiplication by powers of $z$, differentiation,
Laplace transform, and integration. The last property, in particular,
means that the integration of one or a product of two $G$-functions
is a $G$-function. Computer algebra software take advantage of this
property in order to analytically evaluate integrals by representing
the integrand as $G$-function(s).

\paragraph{Elementary properties.}

The following identities can be easily obtained from the definition
(\ref{eq:Meijer-G}),

\begin{subequations} 
\begin{gather}
\begin{aligned}G_{p,q}^{m,n}\left[z\left|{\left(a_{p}\right)\atop \left(b_{q}\right)}\right.\right] & =G_{q,p}^{n,m}\left[z^{-1}\left|{1-\left(b_{q}\right)\atop 1-\left(a_{p}\right)}\right.\right]\\
z^{\sigma}G_{p,q}^{m,n}\left[z\left|{\left(a_{p}\right)\atop \left(b_{q}\right)}\right.\right] & =G_{p,q}^{m,n}\left[z\left|{\left(a_{p}\right)+\sigma\atop \left(b_{q}\right)+\sigma}\right.\right]
\end{aligned}
\label{eq:G-symmetry-translation_formula}\\
G_{p,q}^{m,n}\left[z\left|{\alpha,a_{2},\dots a_{n},a_{n+1},\dots,a_{p}\atop b_{1},\dots,b_{m},b_{m+1},\dots,b_{q-1},\alpha}\right.\right]\hphantom{G_{p,q}^{m,n}}\hphantom{G_{p,q}^{m,n}}\nonumber \\
\hphantom{G_{p-1,q-1}^{m,n-1}}=G_{p-1,q-1}^{m,n-1}\left[z\left|{a_{2},\dots,a_{p}\atop b_{1},\dots,b_{q-1}}\right.\right],\label{eq:G-pars_cancel}
\end{gather}
\end{subequations}where for (\ref{eq:G-pars_cancel}), $n,p,q\geqslant1$.

\paragraph{Integrals containing the $G$-function.}

Amongst the myriad integration formulae available, this work makes
use of 
\begin{multline}
\int_{0}^{\infty}y^{\alpha-1}\left(y+\beta\right)^{-\sigma}G_{p,q}^{m,n}\left[zy\left|{\left(a_{p}\right)\atop \left(b_{q}\right)}\right.\right]dy\\
=\frac{\beta^{\alpha-\sigma}}{\Gamma\left(\sigma\right)}G_{p+1,q+1}^{m+1,n+1}\left[\beta z\left|{1-\alpha,\left(a_{p}\right)\atop \sigma-\alpha,\left(b_{q}\right)}\right.\right].\label{eq:G-Lu75_5.6.7}
\end{multline}

\paragraph{Derivatives.}

The following formulas are employed,\begin{subequations}\label{eq:G-derivatives}
\begin{multline}
\frac{d^{k}}{dz^{k}}\left\{ z^{-b_{1}}G_{p,q}^{m,n}\left[z\left|{\left(a_{p}\right)\atop \left(b_{q}\right)}\right.\right]\right\} =\left(-\right)^{k}z^{-b_{1}-k}\\
\times G_{p,q}^{m,n}\left[z\left|{\left(a_{p}\right)\atop b_{1}+k,b_{2},\dots,b_{q}}\right.\right],\:\left(m\geqslant1\right)
\end{multline}
\vspace*{-\belowdisplayskip}\vspace*{-\abovedisplayskip} 
\begin{align}
z^{k}\frac{d^{k}}{dz^{k}}G_{p,q}^{m,n}\left[z\left|{\left(a_{p}\right)\atop \left(b_{q}\right)}\right.\right] & =G_{p+1,q+1}^{m,n+1}\left[z\left|{0,\left(a_{p}\right)\atop \left(b_{q}\right),k}\right.\right].
\end{align}
\end{subequations}

\paragraph{Representations of special functions.}

If no two of the $b_{h}$ $\left(h=1,\dots,m\right)$ parameters differ
by an integer, all poles of $\Phi\bigl(\cdots;s\bigr)$ in (\ref{eq:Meijer-G})
are simple and then the $G$-function can be expressed as a combination
of the hypergeometric series (\ref{eq:General_hypergeom_series})
as 
\begin{multline}
G_{p,q}^{m,n}\left[z\left|{\left(a_{p}\right)\atop \left(b_{q}\right)}\right.\right]\\
=\sum_{h=1}^{m}\frac{\prod_{j=1}^{m}\Gamma\left(b_{j}-b_{h}\right)^{*}\prod_{j=1}^{n}\Gamma\left(1+b_{h}-a_{j}\right)}{\prod_{j=m+1}^{q}\Gamma\left(1+b_{h}-b_{j}\right)\prod_{j=n+1}^{p}\Gamma\left(a_{j}-b_{h}\right)}\\
\times z^{b_{h}}\pFq p{q-1}\left({1+b_{h}-\left(a_{p}\right)\atop 1+b_{h}-\left(b_{q}\right)^{*}};\left(-\right)^{p-m-n}z\right),\label{eq:G-from_pFq}
\end{multline}
which is valid for $p<q$ or $p=q$ and $\left|z\right|<1$\@. The
notation $\Gamma\left(b_{j}-b_{h}\right)^{*}$ means that this term
is absent when $h=j$.

If any pair of $b_{h}$ parameters differ by an integer, then expression
(\ref{eq:G-from_pFq}) is no longer valid and the singularities in
different terms have to be cancelled out by a limiting process. In
this case the final result will contain a logarithmic singularity
and the $G$-function will then represent a function that is not simply
expandable in a power series. A detailed account of this lengthy process
is given by Ref. \onlinecite{Luke75}\@. Although this case is
relevant in this work, the obtained results can always be expressed
by known functions.

A short list of function representations is given below.\begin{subequations}
\begin{multline}
\pFq pq\left({\left(a_{p}\right)\atop \left(b_{q}\right)};z\right)\\
=\frac{\prod_{j=1}^{q}\Gamma\left(b_{j}\right)}{\prod_{j=1}^{p}\Gamma\left(a_{j}\right)}G_{p,q+1}^{1,p}\left[-z\left|{1-\left(a_{p}\right)\atop 0,1-\left(b_{q}\right)}\right.\right]\label{eq:G-rep:pFq(1)}
\end{multline}
\vspace*{-\belowdisplayskip}\vspace*{-0.5\abovedisplayskip} 
\begin{multline}
J_{\mu}\left(\sqrt{z}\right)J_{\nu}\left(\sqrt{z}\right)\\
=\frac{1}{\sqrt{\pi}}G_{2,4}^{1,2}\left[z\left|{0,\nicefrac{1}{2}\atop \frac{\mu+\nu}{2},-\frac{\mu+\nu}{2},\frac{\mu-\nu}{2},-\frac{\mu-\nu}{2}}\right.\right]\label{eq:G-rep:J_mu-J_nu}
\end{multline}
\vspace*{-\belowdisplayskip}\vspace*{-0.2\abovedisplayskip} 
\begin{align}
J_{\nu}^{2}\left(\sqrt{z}\right) & =\frac{1}{\sqrt{\pi}}G_{1,3}^{1,1}\left[z\left|{\nicefrac{1}{2}\atop \nu,-\nu,0}\right.\right]\label{eq:G-rep:J_nu^2}\\
e^{-z/2}I_{\nu}\left(\frac{z}{2}\right) & =\frac{1}{\sqrt{\pi}}G_{1,2}^{1,1}\left[z\left|{\nicefrac{1}{2}\atop \nu,-\nu}\right.\right]\label{eq:G-rep:H_nu}\\
I_{\nu}\left(\sqrt{z}\right)K_{\nu}\left(\sqrt{z}\right) & =\frac{1}{2\sqrt{\pi}}G_{1,3}^{2,1}\left[z\left|{\nicefrac{1}{2}\atop 0,\nu,-\nu}\right.\right].\label{eq:G-rep:IK}
\end{align}
\end{subequations}

In particular, representation (\ref{eq:G-rep:pFq(1)}) gives meaning
to an hypergeometric function $\pFq pq\left(\cdots;z\right)$ when
$p>q+1$.

\section{Cartesian components of the dielectric tensor\label{sec:DT_Cartesian}}

The components of the dielectric tensor in Cartesian coordinates and
in terms of the Stix parameters are\cite{Brambilla98} 
\begin{align*}
\varepsilon_{xx} & =\hat{S} & \varepsilon_{\left(\substack{xy\\
yx
}
\right)} & =\mp i\hat{D}\\
\varepsilon_{\left(\substack{xz\\
zx
}
\right)} & =N_{\perp}N_{\parallel}\eta & \varepsilon_{yy} & =\hat{S}-N_{\perp}^{2}\hat{\tau}+N_{\perp}^{2}\\
\varepsilon_{\left(\substack{yz\\
zy
}
\right)} & =\pm iN_{\perp}N_{\parallel}\zeta & \varepsilon_{zz} & =\hat{P},
\end{align*}
where $\hat{S}=\frac{1}{2}\bigl(\hat{R}+\hat{L}\bigr)$ and $\hat{D}=\frac{1}{2}\bigl(\hat{R}-\hat{L}\bigr)$
are respectively the $\hat{S}$um and $\hat{D}$ifference (thermal)
Stix parameters, and $\eta=\frac{1}{2}\left(\hat{\mu}+\hat{\nu}\right)-1$
and $\zeta=\frac{1}{2}\left(\hat{\nu}-\hat{\mu}\right)$ are kinetic
parameters.

Then, using the expressions (\ref{eq:Stix_pars_kappa}) for the kappa
Stix parameters, one obtains 
\begin{align*}
\hat{S}_{\kappa} & =1+\sum_{a}\frac{\omega_{pa}^{2}}{\omega^{2}}\sum_{n\to-\infty}^{\infty}\frac{n^{2}}{\mu_{a}}\xi_{0a}\mathcal{Z}_{n,\kappa_{a}}^{\left(\alpha_{a},2\right)}\\
\hat{D}_{\kappa} & =-\sum_{a}\frac{\omega_{pa}^{2}}{\omega^{2}}\sum_{n\to-\infty}^{\infty}n\xi_{0a}\frac{\partial\mathcal{Z}_{n,\kappa_{a}}^{\left(\alpha_{a},2\right)}}{\partial\mu_{a}}\\
\eta_{\kappa} & =-\frac{1}{2}\sum_{a}\frac{\omega_{pa}^{2}}{\omega\Omega_{a}}\frac{w_{a}^{2}}{c^{2}}\sum_{n\to-\infty}^{\infty}\frac{n}{\mu_{a}}\xi_{0a}^{2}\frac{\partial\mathcal{Z}_{n,\kappa_{a}}^{\left(\alpha_{a},1\right)}}{\partial\xi_{na}}\\
\zeta_{\kappa} & =\frac{1}{2}\sum_{a}\frac{\omega_{pa}^{2}}{\omega\Omega_{a}}\frac{w_{a}^{2}}{c^{2}}\sum_{n\to-\infty}^{\infty}\xi_{0a}^{2}\frac{\partial^{2}\mathcal{Z}_{n,\kappa_{a}}^{\left(\alpha_{a},1\right)}}{\partial\xi_{na}\partial\mu_{a}}.
\end{align*}
% \bibliographystyle{aipnum4-1}% \bibliography{revabr,plasma,splasma,matematica,comp,physics},\bibliographystyle{plain}
% \bibliography{revabr,splasma,plasma,matematica,comp,physics}
\bibliography{GaelzerZiebell15_rev}

\end{document}